\documentclass[10pt]{iopart}
\usepackage{url} 
\usepackage{pdflscape} 
\usepackage{amsmath}
\usepackage{amssymb} 
\usepackage{dcolumn}
\usepackage{bm} 
\usepackage{color}
\usepackage{epsfig} 
\usepackage{amsfonts}
\usepackage{graphicx}
\usepackage{subfigure}
\usepackage{dcolumn,pifont}
\usepackage{epsfig}
\usepackage{graphicx}
\usepackage{amsmath}
\usepackage{epsfig}
\usepackage{bm}
\usepackage{times,fancyhdr}
\usepackage{threeparttable}
\usepackage{rotating}

\def\case#1/#2{\textstyle\frac{#1}{#2}}

\newcommand{\be}{\begin{equation}}
\newcommand{\ee}{\end{equation}}
\newcommand{\ben}{\begin{eqnarray}}
\newcommand{\een}{\end{eqnarray}}
\newtheorem{thm}{Theorem}

\newtheorem{rem}{Remark}
\newtheorem{prop}{Proposition}
\newtheorem{defn}{Definition}

\def\be{\begin{equation}}
\def\ee{\end{equation}}
\def\bea{\begin{eqnarray}}
\def\eea{\end{eqnarray}}

\def\case#1/#2{\textstyle\frac{#1}{#2}}

\def\bea{\begin{eqnarray}}
\def\eea{\end{eqnarray}}

\newcommand{\vphi}{\varphi}

\PassOptionsToPackage{linktocpage}{hyperref}
\def\case#1/#2{\textstyle\frac{#1}{#2}}

\begin{document}

\title
{Some remarks about non-minimally coupled scalar field models}

\author{Carlos R. Fadragas}
\address{Department of Physics, Universidad Central de
Las Villas, Santa Clara CP 54830, Cuba}
\ead{fadragas@uclv.edu.cu}

\author{Genly Leon}

\address{Instituto de F\'{\i}sica, Pontificia Universidad  Cat\'olica
de Valpara\'{\i}so, Casilla 4950, Valpara\'{\i}so, Chile}
\ead{genly.leon@ucv.cl}

\begin{abstract}
Several results related to flat Friedmann-Lema\^{\i}tre-Robertson-Walker models in the conformal (Einstein) frame of scalar–tensor gravity theories are extended. Scalar fields with arbitrary (positive) potentials and arbitrary coupling functions are considered. Mild assumptions under such functions (differentiable class, number of singular points, asymptotes, etc) are introduced in a straightforward manner in order to characterize the asymptotic structure on a phase space. We pay special attention to the possible scaling solutions. Numerical evidence confirming our results is presented.
\end{abstract}

\pacs{98.80.-k, 98.80.Jk,98.80.Cq., 95.36.+x,  95.30.Sf, 04.20.Ha}
\date{\today}

\maketitle


\section{Introduction}

Recent astrophysical observations suggest that the universe is
currently experiencing an accelerated expansion
\cite{Komatsu:2010fb,Kowalski:2008ez,Allen:2007ue,Abazajian:2008wr,Jarosik:2010iu,Perlmutter:1998np,Bennett:2003bz,Tegmark:2003ud,Allen:2004cd,Tegmark:2003uf,Komatsu:2003fd,Hinshaw:2003ex,Spergel:2003cb,Peiris:2003ff,Riess:1998cb,Riess:2004nr,Jassal:2004ej,Freedman:2000cf,Mould:1999ap,Choudhury:2003tj,Padmanabhan:2002vv}.  To explain this feature of the universe one choice is
to introduce the concept of
Dark Energy (DE)
(see \cite{Sahni:1999gb,Peebles:2002gy,Copeland:2006wr} and references
therein), which could be a cosmological constant, a quintessence field 
\cite{Ratra:1987rm,Wetterich:1987fm,Liddle:1998xm,Kolda:1998wq,Sahni:2002kh,Padmanabhan:2002ji,Dutta:2009yb}, a phantom
field \cite{Caldwell:1999ew,Boisseau:2000pr,Nojiri:2003vn,Onemli:2004mb,Aref'eva:2004vw,Saridakis:2008fy},  the quintom field
\cite{Feng:2004ad,Guo:2004fq,Zhang:2005eg,Aref'eva:2005fu,Zhao:2006mp,Lazkoz:2006pa,Vernov:2006dm,Lazkoz:2007mx,Setare:2008pz,Cai:2009zp,Aref'eva:2009xr,Leon:2012vt}, among other examples. Another choice is to consider Higher Order Gravity (HOG) theories, say the $f(R)$- models (see
\cite{Carroll:2004de,Nojiri:2006ri,Capozziello:2007ec,Sotiriou:2008rp,
Nojiri:2010wj,DeFelice:2010aj,Motohashi:2010zz,Capozziello:2011et,Giambo:2008sa} and references
therein). Other modified gravitational scenarios are the extended nonlinear massive gravity scenario
\cite{Cai:2013lqa,Gannouji:2013rwa,Leon:2013qh,Cai:2012ag}, the
Teleparalell Dark Energy (TDE) model \cite{Geng:2011aj,Geng:2011ka,Xu:2012jf} and some generalizations of the TDE \cite{Harko:2014sja,Kofinas:2014aka}. All of these scenarios have very interesting cosmological features. 

Another interesting effective scalar-field model, which is related to inhomogeneous cosmologies, is the so-called ``morphon'' field that arises as an effective scalar field model for averaged cosmologies \cite{Buchert:2006ya,Roy:2011za}. In this case the scalar field is  not interpreted as a source of the Einstein equations, but as a mean field description of averaged inhomogeneities. In this case the ``backreaction'' effects (due to averaged expansion and shear fluctuations, the averaged 3-Ricci curvature, averaged pressure gradients and frame fluctuation terms) is formally equivalent to the dynamics of a homogeneous, minimally coupled scalar field. This relation was completely addressed in \cite{Buchert:2006ya}.

Although our results are more widely applicable if the scope is widened to effective scalar fields like the morphon field, in this paper the DE contribution is modelled as a conventional self-interacting quintessence scalar field, $\phi,$ with
potential $V(\phi)$ \cite{Ratra:1987rm,Wetterich:1987fm,Liddle:1998xm,Kolda:1998wq,Sahni:2002kh,Padmanabhan:2002ji,Dutta:2009yb}. 

In the inflationary universe scenarios the
potential $V(\phi)$ must satisfy some requirements which are necessary to
lead to the early-time acceleration of the expansion
\cite{Billyard:1999ct,Copeland:2004cz,Copeland:1993ie,Kolb:1995iv,Lidsey:1995np}. Exponential (de Sitter) expansion arises, for example, as a result of considering a constant potential $V (\phi) = V_0$, whereas the power-law inflationary solutions arise when considering an exponential potential $V (\phi) = V_0\exp(-\lambda\phi)$ \cite{Lucchin:1984yf,Burd:1988ss}.  In the reference \cite{Alho:2014fha} have been investigated a minimally coupled  scalar field evolving in the quadratic potential $V(\phi)=\frac{1}{2}m^2 \phi^2$ in the flat Friedmann-Lema\^{i}tre-Robertson-Walker (FLRW) metric. There  was provided a global picture of the solution space by means of a regular global dynamical system defined in extended compact space. This system is suitable for obtaining global piecewise approximations for the late-time attractor solution due to the large range of convergence for center manifold expansions and the associated approximants as compared with the slow-roll approximation and associated slow-roll approximants \cite{Alho:2014fha}.

Several gravity theories consider
multiple scalar fields with exponential potential, e.g., assisted inflation scenarios
\cite{Chimento:1998aq,Guo:2003eu,Coley:1999mj,Copeland:1999cs,Hartong:2006rt,Malik:1998gy},
quintom dark energy paradigm
\cite{Guo:2004fq,Zhang:2005eg,Zhao:2006mp,Lazkoz:2006pa,Lazkoz:2007mx} and
others.  The potential have been considered as positive and negative
exponential \cite{Heard:2002dr}, single and
double exponential
\cite{Copeland:1997et,Sen:2001xu,Jarv:2004uk,Li:2005ay,Huang:2006ku,Gonzalez:2007hw,Tzanni:2014eja,Arias:2003fd,Cardenas:2002hw,Huey:2004qv,Gonzalez:2005ie,Zhang:2005rj,Capozziello:2005ra,Gonzalez:2006cj,Carvalho:2006fy,Boehmer:2008av,Barrow:2008ju,Chen:2008pz,Basilakos:2011rx,Tsamparlis:2011cw,Farajollahi:2012nr,Cicoli:2012tz}, etc. Multiple scalar fields can be found at
\cite{vandenHoogen:2000cf,Bagrov:1990pf,Damour:1992we,EspositoFarese:1992nd,Rainer:1996gw,Wang:1998mm,Rodrigues:2011zi,Tsujikawa:2006mw,Galli:2010qx,Clesse:2011jt,Xue:2012wi,Alho:2013vva}.

Very interesting cases are non-minimally coupled scalar fields that appear in the string theory context \cite{Gasperini:2007zz} or in the Scalar-tensor
theories (STT) context
\cite{Brans:1961sx,Wagoner:1970vr,O'Hanlon:1972my,O'Hanlon:1972hq,Bekenstein:1977rb,Bergmann:1968ve,Nordtvedt:1970uv,Fujii:2003pa,Maeda:2009js,Faraoni:2004pi}. Coupled quintessence models were investigated by means of phase-space studies for example in  \cite{Huey:2004qv,Gonzalez:2005ie,Zhang:2005rj,Gonzalez:2006cj,Boehmer:2008av,Chen:2008pz,Amendola:1999er,TocchiniValentini:2001ty,Billyard:2000bh}. Specific non-minimally coupled subclass of Horndeski scalar-tensor theories arising from the decoupling limit of massive gravity by covariantization were studied in \cite{deRham:2011by,Heisenberg:2014kea}. 
In the reference \cite{Skugoreva:2014gka}, it was investigated a flat FLRW scalar field with potential of types $V(\phi)=\phi^n$ and $V(\phi)=\phi^{n_1}+\phi^{n_2}$, conformally coupled to the Ricci scalar, $R$, through the function $-\xi B(\phi) R$, where $\xi$ is the coupling constant and $B(\phi)=\phi^N.$ The authors worked in the Jordan frame in the absence of matter. It was presented there a global picture of the phase space by means of compact variables. Some exact solutions, for some choices of the slopes of the potential and the coupling function, were discussed there. 
In the reference \cite{Aref'eva:2012au}, it was investigated a scalar field non-minimally coupled to the Ricci scalar evolving in
Higgs-like (quadratic) potentials  plus a negative cosmological constant.  Double exponential potential and exponential coupling function were discussed in \cite{Tzanni:2014eja} as well as in \cite{Arias:2003fd,Cardenas:2002hw,Gonzalez:2005ie,Gonzalez:2006cj}, under the ansatz $\dot\phi=\lambda H$.

 In the reference \cite{Fadragas:2013ina}, it was performed a detailed dynamical analysis of Kantowski-Sachs, Locally Rotationally Symmetric
(LRS) Bianchi I and LRS Bianchi III models by means of the method
of $f$-devisers presented in \cite{Escobar:2013js} and applied in \cite{delCampo:2013vka} to scalar field cosmologies in the framework of a generalized Chaplygin gas. This method, that allows us to perform the whole analysis for a wide range of potentials, is a modification of the method first introduced in \cite{Fang:2008fw}. The original method was used for investigating flat FLRW scalar
field cosmologies \cite{Dutta:2009yb,Fang:2008fw,Leyva:2009zz,Matos:2009hf,Copeland:2009be,UrenaLopez:2011ur} and it was generalized to several cosmological contexts in \cite{Escobar:2011cz,Escobar:2012cq,Xiao:2011nh,Farajollahi:2011ym}. As a drawback of the method of $f$-devisers it cannot be applied to some specific inflationary potentials like the logarithmic $V(\phi)\propto \phi ^p \ln ^q(\phi )$ \cite{Barrow:1995xb} and the generalized exponential one $V(\phi)\propto \phi^n \exp\left(-\lambda\phi^m\right)$ \cite{Parsons:1995ew} since the resulting $f$-functions are not single valued, then one should apply asymptotic techniques in order to extract the dominant branch at large $\phi$-values as in \cite{Barrow:1995xb,Parsons:1995ew}.

The idea to obtain general results for scalar field cosmologies only providing general features of the potentials and coupling functions, is not new. Preceding works for a large variety of non-negative potentials are
\cite{Rendall:2004ic,Rendall:2006cq,Rendall:2005if,Rendall:2005fv,Foster:1998sk,Miritzis:2003ym,Miritzis:2005hg,Hertog:2006rr}. 
In the reference \cite{Hertog:2006rr}, it have been extended many of the
results obtained in \cite{Miritzis:2003ym} by considering arbitrary
potentials. In \cite{Foster:1998sk} it has been shown that for a
large class of flat FLRW cosmologies with scalar fields with arbitrary
potential, the past attractor corresponds to exactly integrable cosmologies with
a massless scalar field. A list of integrable models with a minimally coupled scalar field, including double exponential potentials, in the absence of matter, is presented in  \cite{Fre:2013vza}. Integrable non-minimally coupled scalar field models in the Jordan frame were investigated in \cite{Kamenshchik:2013dga}. In \cite{Leon:2008de} were investigated flat FLRW cosmologies based on STTs. The new asymptotic expansions  for the cosmological solutions near the initial space-time singularity obtained there contains as particular cases those studied in \cite{Foster:1998sk}. The proof of the local initial singularity theorem presented in \cite{Leon:2008de} have been improved in \cite{Lap_Lambert}. 
Additionally in \cite{Lap_Lambert} were presented several results corresponding the late-time dynamics for the case of a scalar field non-minimally coupled to dark matter. In this paper we extent these results by adding a radiation fluid. This leads to a more realistic cosmological model in the framework of the so-called \emph{complete cosmological dynamics} \cite{Avelino:2013wea}, that is, a viable cosmological model should describe a radiation dominated era (RDE) before entering a matter dominated era (MDE), which should be succeeded by the current late-time accelerated expansion \cite{Leon:2012vt,Avelino:2013wea,Cruz:2014iva}. The transition from one era to the next can be understood, in the language of dynamical systems, in terms of the so-called heteroclinic sequences \cite{WE,Coley:2003mj,Heinzle:2007kv,UrenaLopez:2011ur}.

Models arising in the  conformal frame of $F(R)$ theories were considered. e.g., in \cite{Giambo':2009cc,Miritzis:2010cr,Nova_Science,Miritzis:2011zz,Lap_Lambert}. In \cite{Giambo':2009cc} were investigated flat and negatively
curved FLRW models with a perfect
fluid matter source and a nonminimally coupled scalar field $\phi$ with potential $V(\phi)$ (related to the $F(R)$ function). There was proved that for potentials that eventually becomes non-negative as $\phi\rightarrow \pm \infty$ and with a finite number of critical points, the non-negative local minima and the horizontal asymptotes approached
from above by it, are asymptotically stable. For a nondegenerated minimum with zero critical value and $\gamma>1$, there is a transfer of energy from the fluid to the scalar field, which eventually dominates the expansion in  a generic way. In \cite{Nova_Science} it was developed a mathematical procedure for investigate the dynamics when $|\phi|\rightarrow +\infty$ of non-minimally coupled scalar fields models interacting with Dark Matter in the presence of radiation. There were studied the modified gravity models $f(R)=R+\alpha R^2$ (quadratic gravity) and $F(R)=R^n$ in the STT frame. For quadratic gravity, the equilibrium point corresponding to de Sitter solution is locally asymptotically unstable (saddle point).

The aim of the paper is to extent several results in
\cite{Foster:1998sk,Miritzis:2003ym,Leon:2008de,Giambo':2009cc, Miritzis:2010cr,Miritzis:2011zz,Tzanni:2014eja} for the general case of arbitrary potentials and arbitrary couplings. It is described the early and late-time dynamics of the models and we pay special attention to the possible scaling solutions. We follow the method first introduced in \cite{Foster:1998sk} and extended in \cite{Nova_Science} for the analysis of the limit $\phi\rightarrow +\infty$. We examine the example of a double exponential potential and our new results complement those in \cite{Tzanni:2014eja}. Additionally,  we revisit the example of a powerlaw coupling function and an Albrecht-Skordis potential, first introduced in \cite{Leon:2008de}, and then extended in section 4.4 of \cite{Nova_Science}.

\section{Basic framework}

The action for a general class of STT, written in the so-called
Einstein frame (EF), is given by \cite{Kaloper:1997sh}:
\begin{align}&\int  d{ }^4 x \sqrt{|g|}\left\{\frac{1}{2} R-\frac{1}{2} g^{\mu
\nu}\nabla_\mu\phi\nabla_\nu\phi-V(\phi)+\chi(\phi)^{-2}
\mathcal{L}(\mu,\nabla\mu,\chi(\phi)^{-1}g_{\alpha\beta})\right\}.\label{eq1}
\end{align} We use a system of units in which $8\pi G=c=\hbar=1.$
In this equation $R$ is the curvature scalar, $\phi$ is the a
scalar field \footnote{For a discussion about the
regularity of the conformal transformation, or the equivalence
issue of the two frames, see for example
\cite{Magnano:1990qu,Magnano:1993bd,Cotsakis:1993vm,Teyssandier:1995wr,Schmidt:1995ws,Cotsakis:1995wt,Capozziello:1996xg,Faraoni:1998qx,Faraoni:2006fx,Faraoni:2007yn,Romero:2012hs,Quiros:2011wb}
and references therein.},  $\nabla_\alpha$
is the covariant derivative, $V(\phi)$ is the quintessence self-interaction potential,
$\chi(\phi)^{-2}$ is the coupling function, $\mathcal{L}$
is the matter Lagrangian, $\mu$ is a collective name for the
matter degrees of freedom. 

The matter energy-momentum tensor  is defined by
\begin{equation}T_{\alpha
\beta}=-\frac{2}{\sqrt{|g|}}\frac{\delta}{\delta g^{\alpha
\beta}}\left\{\sqrt{|g|}
 \chi^{-2}\mathcal{L}(\mu,\nabla\mu,\chi^{-1}g_{\alpha
 \beta})\right\}.\label{Tab}\end{equation}

Let's define 
$$Q_\beta\equiv\nabla^\alpha T_{\alpha \beta}=-\frac{1}{2}T\frac{1}{\chi(\phi)}\frac{\mathrm{d}\chi(\phi)}{\mathrm{d}\phi}\nabla_{\beta}\phi,\;
 T=T^\alpha_\alpha.$$
Since there is an exchange of energy between the scalar and the
background fluids, the energy is not separately conserved for each
component. Instead, the continuity equation for each fluid reads \cite{Curbelo:2005dh}:
\begin{equation}
\dot\rho_m+3H(\rho_m +p_m)=Q, \label{backconteq}
\end{equation}
\begin{equation}
\dot\rho_{DE}+3H(\rho_{DE}+p_{DE})=-Q, \label{phconteq}
\end{equation}
where the dot accounts for derivative with respect to the cosmic
time and $Q$ is the interaction term. Now, defining the total energy density and the total pressure as  $\rho_T=\rho_m+\rho_{DE}$ and $p_T= p_m+p_{DE}$, respectively, then the total energy density is indeed
conserved in the sense $\dot\rho_T+3H(\rho_T+p_T)=0$. To specify the general
form of the interaction term we can look at a scalar-tensor
theory of gravity \eqref{eq1} and the
interaction term $Q$ in equations (\ref{backconteq}) and
(\ref{phconteq}), can be written in the following form:
\begin{equation}
Q=-\frac{1}{2}(4-3\gamma)\rho_m H\left[a\frac{d \ln{\chi}}{da}\right],
\label{interactionterm}
\end{equation}
where we have assumed that the coupling $\chi$ can be written as a function of the
scale factor  through $\chi(\phi(a))$. Comparing this with other
interaction terms in the bibliography, one can obtain the
functional form of the coupling function $\chi$ at each case.
In the reference \cite{Guo:2004vg}, for instance,
$Q=3Hc^2(\rho_{DE}+\rho_m)= 3c^2H\rho_m(r+1)/r$, where $c^2$
denotes the transfer strength and $r\equiv
\Omega_{m}/\Omega_{DE}$. If one compares this expression with
(\ref{interactionterm}) one obtains the following coupling
function:
\begin{equation}
\chi(a)=\chi_0\exp\left[-\frac{6}{4-3\gamma}{\int\frac{d a}{a}\left(\frac{r+1}{r}\right)c^2}\right],
\label{xizhang}
\end{equation}
where $\chi_0$ is an arbitrary integration constant. If
$c^2=c_0^2=const.$ and $r=r_0=const.$, then
$\chi=\chi_0\;a^{\frac{6 c_0^2(r_0+1)}{(4-3\gamma)r_0}}$. It is well-known that a suitable coupling can
produce scaling solutions, although the way to fix the coupling is
not univocally determined. In reference
\cite{Amendola:1999er,Amendola:1999qq}, for instance, the coupling
is introduced by hand. In
\cite{Chimento:2003iea,Chimento:2000kq,Chimento:2003sb} the form of
the interaction term is fixed by the requirement that the ratio of
the energy densities of DM and quintessence has an stable fixed
point during the evolution that solves the coincidence; in
\cite{Chimento:2003iea} a suitable interaction between the
quintessence field and DM leads to a transition from the
domination matter era to an accelerated expansion epoch for the
model proposed in
\cite{Chimento:2003iea,Chimento:2000kq,Chimento:2003sb}. In \cite{Bean:2000zm} the coupling function is chosen as a Fourier expansion
around some minimum of the (dilaton) scalar field.

It is well known that the HOG theories \cite{Carroll:2004de,Nojiri:2006ri,Capozziello:2007ec,Sotiriou:2008rp,
Nojiri:2010wj,DeFelice:2010aj,Motohashi:2010zz,Capozziello:2011et,Giambo:2008sa} derived from the action
\begin{equation}
S=\int d^{4}x\sqrt{-g}\left\{\frac{1}{2}F\left(R\right)+\mathcal{L}(\mu,\nabla\mu,g_{\alpha \beta})\right\},
\label{action_F(R)}
\end{equation} and the STT with action \begin{equation}
\widetilde{S}=\int d^{4}x\sqrt{-\widetilde{g}}\left\{\frac{1}{2}
\widetilde{R}-\left(  \nabla\phi\right)  ^{2}-V\left( \phi\right)
 +e^{-2\sqrt{2/3}\phi}\mathcal{L}\left(\mu,\nabla\mu,
e^{-\sqrt{2/3}\phi
}\widetilde{g}_{\alpha \beta}\right)  \right\},\label{action}
\end{equation}  are conformally equivalent under the transformation, 
\begin{subequations}
\begin{align}
&\widetilde{g}_{\mu\nu}=F'(R)  g_{\mu\nu}, \\
&\phi=\sqrt{\frac{3}{2}}\ln F'(R), \label{sfield}\\
&V\left(R(\phi)\right)  =\frac{1}{2\left(F'(R)\right)
^{2}}\left(  R F'(R)-F(R)\right), \label{potential}
\end{align}
\end{subequations}
where it is assumed that (\ref{sfield}) can be solved for $R$ to obtain a
function $R\left(\phi\right)$, in order to get the potential \eqref{potential} as an explicit function of $\phi$. 
It is easy to note that the model arising from the action
\eqref{action} can be obtained from \eqref{eq1} with the choice
$\chi(\phi)=e^{\sqrt{2/3}\phi}.$ 
Thus, the results in \cite{Leon:2008de} and in \cite{Giambo':2009cc} can be obtained as particular cases by investigating a general
class of models containing both STTs and $F(R)$ gravity.

Using the above approach, we obtain that the quadratic gravity model $F(R)=R+\alpha R^2$ is conformally equivalent to a non-minimally coupled scalar field with the potential
$V\left( \phi\right) =\frac{1}{8\alpha}\left(
1-e^{-\sqrt{2/3}\phi}\right) ^{2}, \; \alpha>0.$ This potential has only one local minimum at $\phi=0$, and the asymptote $V_{\infty}=\frac{1}{8 \alpha}$ which is approached by below by the potential as $\phi\rightarrow +\infty.$  The zero minimum at $\phi=0$ of this potential is stable, but it cannot provide the mechanism for the late-time acceleration since $V$ and $H$ asymptotically approach to zero \cite{Miritzis:2003ym,Miritzis:2005hg,Macnay:2008nw}. On the other hand, concerning the upper asymptote of $V$ as $\phi\rightarrow +\infty$, it follows that the potential has exponential order zero (see definition \ref{WBI}) as $\phi\rightarrow +\infty.$ Thus, using Proposition 6 of \cite{Nova_Science}, follows that the de Sitter configuration $\phi\rightarrow +\infty, V(\phi)=V_{\infty}, H(\phi)=\sqrt{\frac{V_{\infty}}{3}}$ is unstable to perturbations along the $\phi$-axis in a neighborhood of ``infinity'', and thus it cannot represent the late-time solution.  

In the reference \cite{Macnay:2008nw} were imposed conditions on the
function $F\left( R\right)$ with corresponding potential
(\ref{potential}) using the restrictions on $V(\phi)$ obtained in the papers
\cite{Rendall:2004ic,Rendall:2006cq,Rendall:2005if,Rendall:2005fv}. There it was exploited from the mathematical view point the connection between $F(R)$-gravity and nonminimally coupled scalar fields and were proved mathematically rigorous results. In \cite{Leon:2013bra}, it was investigated a generic class of $f(R)$ models for the Kantowski-Sachs metric using dynamical systems tools.  

\subsection{The Field Equations}

In this section we investigate the action \eqref{eq1} for the flat FLRW  metric:
\begin{align}d s^2 = -d t^2 +a(t)^2\left({d r^2}+r^2\left(d \theta^2+\sin^2\theta d \varphi^2\right)\right).
\label{flatFRW}
\end{align} where $a(t)$ is the scale factor. The Hubble expansion scalar is  $H=\dot a/a$, where the dot means derivative with respect time.  
We assume that the energy-momentum tensor \eqref{Tab} is $T^\alpha_\beta=\text{diag} \left(-\rho_m,p_m,p_m,p_m\right),$ where $\rho_m$ and
$p_m$ are respectively the isotropic energy density and the
isotropic pressure of barotopic matter with equation of state $p_m=(\gamma-1)\rho_m.$
We include a radiation source with energy density $\rho_r$ since we want to
investigate the possible scaling solutions in the radiation
regime. We neglect ordinary (uncoupled) barotropic matter.

The cosmological equations with the above ``ingredients'' are
\begin{subequations}\label{Syst:12}
\ben \dot H &=& -\frac{1}{2}\left(\gamma
\rho_m
+\frac{4}{3}\rho_r+\dot\phi^2\right),\label{Raych}\\
\dot \rho_m&=& -3\gamma H \rho_m-\frac{1}{2}(4-3\gamma)\rho_m\dot\phi
\frac{\mathrm{d}\ln
\chi(\phi)}{\mathrm{d}\phi},\label{consm}\\
\dot \rho_r &=& -4 H \rho_r,\label{consr}\\
\ddot \phi&=& -3 H \dot \phi-\frac{\mathrm{d}
V(\phi)}{\mathrm{d}\phi}+\frac{1}{2}(4-3\gamma)\rho_m
\frac{\mathrm{d}\ln
\chi(\phi)}{\mathrm{d}\phi}\label{KG},\\
3H^2 &=& \frac{1}{2}\dot\phi^2+V(\phi)+\rho_m+\rho_r. \label{Fried}
\een
\end{subequations} 
We assume the general
hypothesis $V(\phi)\in C^3, V(\phi)>0, \chi(\phi)\in C^3$ and
$\chi(\phi)>0$ and thus the dynamical
system \eqref{Syst:12} is of class $C^2.$ In order to derive our results we shall consider further
assumptions which shall be clearly stated when necessary.  We
assume $\rho_m,\rho_r\geq 0$, and $0<\gamma<2, \gamma\neq
\frac{4}{3}.$ The later assumption excludes the
possibility that the background matter behaves as radiation which in rigor is automatically decoupled from the scalar field (since the energy-momentum tensor for radiation is traceless).

\section{Dynamical systems analysis}

In the following, we study the late time behavior of solutions of
\eqref{Syst:12}, which are
expanding at some initial time, i. e., $H(0)>0.$ The state vector
of the system is $\left(\phi, \dot\phi, \rho_m, \rho_r, H\right).$
Defining $y:=\dot\phi,$ we rewrite the autonomous system as 
\begin{subequations}\label{Syst:13}
\ben
\dot H &=& -\frac{1}{2}\left(\gamma \rho_m
+\frac{4}{3}\rho_r+y^2\right),\label{newRaych} \\
\dot \rho_m&=& -3\gamma H \rho_m-\frac{1}{2}(4-3\gamma)\rho_m y
\frac{\mathrm{d}\ln
\chi(\phi)}{\mathrm{d}\phi},\label{newconsm}\\
\dot \rho_r &=& -4 H \rho_r,\label{newconsr}\\
\dot y&=& -3 H y-\frac{\mathrm{d}
V(\phi)}{\mathrm{d}\phi}+\frac{1}{2}(4-3\gamma)\rho_m
\frac{\mathrm{d}\ln
\chi(\phi)}{\mathrm{d}\phi}\label{newKG},\\
\dot\phi&=&y,\label{dotphi}
\een
\end{subequations} subject to the
constraint \be 3H^2 = \frac{1}{2}y^2+V(\phi)+\rho_m+\rho_r.
\label{newFried0}\ee

\begin{rem}\label{decreasing}
Using standard arguments of ordinary differential equations
theory, follows from equations \eqref{newconsm} and \eqref{newconsr}
that the signs of $\rho_m$ and $\rho_r,$
respectively, are invariant. This means that if $\rho_m>0$ and $\rho_r>0$
for some initial time $t_0,$ then $\rho_m(t)>0,$ and $\rho_r(t)>0$ throughout the solution.
From \eqref{newRaych} and \eqref{newFried0} and only if additional conditions are
assumed, for example $V(\phi)\geq 0$ and $V(\phi_*)= 0$ for some $\phi_*$, follows that the sign of $H$ is invariant.
From \eqref{newconsr} and \eqref{newRaych} follows that $\rho_r$ and
$H$ decreases. Also, defining $\epsilon=\frac{1}{2} y^2+V(\phi),$
follows from \eqref{newconsm}, \eqref{newKG} and \eqref{newconsr} that
\be\dot\epsilon+\dot\rho_m+\dot\rho_r=-3 H(y^2+\gamma \rho_m+\frac{4}{3}\rho_r)\label{decr}.\ee
Thus, the total energy density contained in the dark sector is
decreasing.
\end{rem}

The system \eqref{Syst:13} defines a dynamical system in
the phase space
\begin{equation}\Omega=\{\left(H,\rho_m, \rho_r,y,\phi\right)\in\mathbb{R}^4|3H^2 =
\frac{1}{2}y^2+V(\phi)+\rho_m+\rho_r\}\label{newFried}.\end{equation}

Let's assume in first place that the potential function has a local minimum
$V(0)=0.$ This implies that the point $(0,0,0,0,0)$ is a singular
point of \eqref{Syst:13} which implies that an initially expanding universe ($H>0$) should expand
forever. Indeed, the set  $\left\{(H, \rho_m, \rho_r, y, \phi)\in\Omega| H=0
\right\}$ is invariant under the flow of \eqref{Syst:13}. 
Besides, the sign of $H$ is invariant. Otherwise, if the sign of $H$ changes,
a trajectory with $H(0)>0$ can passing through $(0,0,0,0,0),$
violating the existence an uniqueness theorem for ODEs.

The proposition 2 of \cite{Miritzis:2003ym} can be generalized to
this context as follows.

\begin{prop}\label{Proposition I} Suppose that $V\geq 0$ and
$V(\phi)=0\Leftrightarrow \phi=0.$ Let $A$ such that $V$ bounded
in $A$ implies $ V'(\phi)$ is bounded in $A.$ If there exists a
constant $K,$ $K\neq 0$ such that
$$\chi'(\phi)/\chi(\phi)\leq 2 K/(2-\gamma)(4-3\gamma).$$
Then, $$\lim_{t\rightarrow\infty}(\rho_m, \rho_r,y)=(0,0,0).$$
\end{prop}

{\bf Proof}. Consider the trajectory passing through an arbitrary
point $(H, \rho_m, \rho_r, y, \phi)\in\Omega $ with $H>0$ at $t=t_0.$ Since
$H$ is positive and decreasing we have that $\lim_{t\rightarrow
\infty} H(t)$ exists and it is a nonnegative number $\eta$;
besides, $H(t)\leq H(t_0)$ for all $t\geq t_0.$ Then, from \eqref{newFried} follows that each term $\rho_m,$ $\rho_r$, $1/2 y^2,$
and $V(\phi)$ is bounded by $3H(t_0)^2$ for all $t\geq t_0.$

Let defined $A=\left\{\phi: V(\phi)\leq 3H(t_0)^2\right\}.$ Then, the
trajectory is such that $\phi$ remains in the interior of $A$ and additionally $V'(\phi)$ is bounded for $\phi\in A.$

From equation \eqref{newRaych} follows
$$-\int_{t_0}^{t}\left(\frac{1}{2}y^2+\frac{\gamma}{2}\rho_m+\frac{2}{3}\rho_r\right)\mathrm{d}
t = H(t)-H(t_0).$$ Taking the limit $t\rightarrow\infty,$ we
obtain
\begin{equation}\frac{1}{2}\int_{t_0}^{\infty}\left(y^2+\gamma\rho_m+\frac{4}{3}\rho_r\right) d t = H(t_0)- \eta \implies
\int_{t_0}^{\infty}\left(y^2+\gamma\rho_m+\frac{4}{3}\rho_r\right) \mathrm{d} t
<\infty.\label{eq10}
\end{equation}

Taking the time derivative of $f(t)= y^2+\gamma\rho_m+\frac{4}{3}\rho_r$ and making
use of the hypothesis for $\chi(\phi)$ we obtain
\begin{align}&\frac{d}{d t}\left(y^2+\gamma\rho_m+\frac{4}{3}\rho_r\right)\leq
y\left(-2 V'(\phi) +K\rho_m\right)-\frac{16}{3}\rho_r H.\nonumber\end{align} As we have
seen, $y$, $\rho_m$, $\rho_r$ and $H$ are bounded for $t\geq t_0$, and by the hypothesis for
$V(\phi)$, $ V'(\phi)$ is bounded. From this facts follow that the
time derivative of $f$ is bounded. Since $f$ is a nonnegative
function, the convergence of $\int_{t_0}^{\infty} f(t) \mathrm{d}
t$ implies $\lim_{t\rightarrow\infty}f(t)=0.$ Hence, we have that
$$\lim_{t\rightarrow\infty} (\rho_m, \rho_r, y)=(0,0,0). \blacksquare$$

The hypotheses in \ref{Proposition I} concerning to the scalar field self-interacting potential are not very restrictive \cite{Miritzis:2003ym}. The hypothesis for $\chi(\phi)$ is satisfied by a large class of coupling functions too, including the exponential ones. 

Under the same hypothesis of proposition \ref{Proposition I}, we
can generalize the proposition 3 in \cite{Miritzis:2003ym}.
\begin{prop}\label{Proposition II} Suppose that $V'(\phi)>0$ for
$\phi>0$ and $V'(\phi)<0$ for $\phi<0.$ Then, under the same
hypotheses as in proposition \ref{Proposition I},
$\lim_{t\rightarrow\infty}\phi$ exists and is equal to $+\infty$,
$0$ or $-\infty.$
\end{prop}

{\bf Proof}. Using the same argument as in Proposition
\ref{Proposition I}, $\exists \lim_{t\rightarrow\infty}
H(t)=\eta.$ If $\eta=0,$ then by the restriction \eqref{newFried} we
obtain $\lim_{t\rightarrow\infty} V(\phi(t))=0.$ Since $V$ is
continuous and $V(\phi)=0\Leftrightarrow \phi=0$ this implies that
$\lim_{t\rightarrow\infty} \phi(t)=0.$

Suppose that $\eta>0.$ From \eqref{newFried} we obtain that
$\lim_{t\rightarrow\infty} V(\phi(t))=3\eta^2.$ Therefore, exists
$t'$ such that $V(\phi)>3\eta^2/2$ for all $t>t'.$ From this fact
follows that $\phi$ cannot be zero for some $t>t'$ because $\phi=0
\Leftrightarrow V(\phi)=0.$ Then, the sign of $\phi$ is invariant
for all $t>t'.$

Suppose that $\phi$ is positive for all $t>t'.$ Since $V$ is an
increasing function of $\phi$ in $(0,+\infty),$ we have that
$\lim_{t\rightarrow\infty} V(\phi(t))=3\eta^2\leq
\lim_{\phi\rightarrow\infty} V(\phi).$  By the continuity and
monotony of $V$ it is obvious that the equality holds if, and only
if, $\lim_{t\rightarrow\infty} \phi(t)=+\infty.$

If $\lim_{t\rightarrow\infty} V(\phi(t))<
\lim_{\phi\rightarrow\infty} V(\phi),$ then there exists
$\bar{\phi}\geq 0$ such that $$\lim_{t\rightarrow\infty}
V(\phi(t))=V(\bar{\phi}).$$ Since $V$ is continuous and strictly
increasing we have that
$$\lim_{t\rightarrow\infty} \phi=\bar{\phi}.$$

By proposition \ref{Proposition I}, $\lim_{t\rightarrow\infty}
(\rho_m(t), \rho_r(t), y(t))=(0,0,0)$. Besides, $H$ and
$\chi'(\phi)/\chi(\phi)$ are bounded. Therefore, taking the limit
as $t\rightarrow\infty$ in \eqref{newKG} we find that
$$\lim_{t\rightarrow\infty}\frac{d}{d
t}y=-V'(\bar{\phi})<0.$$ Hence, there exists $t''> t'$ such that
$\frac{d}{d t}y<-V'(\bar{\phi})/2$ for all $t\geq t''.$ This
implies $$y(t)- y(t'')=\int_{t''}^{t}\left(\frac{d}{d t}y\right)
dt<-\frac{ V'(\bar{\phi})}{2}(t- t''),$$ that is, $y(t)$ takes
negative values with arbitrary large modulus as $t$ increases,
which is not possible since $\lim_{t\rightarrow\infty}y(t)=0.$

Hence, if $\phi>0$ for all $t>t',$ we have that
$\lim_{t\rightarrow\infty}\phi=+\infty.$ Similarly, when $\phi<0$
for all $t>t',$ we have $\lim_{t\rightarrow\infty}\phi=-\infty.$
$\blacksquare$

From this we conclude that, if initially
$3H(t_0)^2<\min\left\{\lim_{\phi\rightarrow\infty}V(\phi),
\lim_{\phi\rightarrow -\infty}V(\phi)\right\},$  then,
$\lim_{t\rightarrow\infty}H(t)=0.$ Indeed, we have that
$\lim_{t\rightarrow\infty}\phi$ is equal to $+\infty$, $0$ or
$-\infty.$ If $\lim_{t\rightarrow\infty}\phi=+\infty$, then from
the restriction \eqref{newFried}, follows
$$3 \eta^2=\lim_{t\rightarrow\infty}V(\phi(t))=
\lim_{\phi\rightarrow\infty} V(\phi)>3 H(t_0)^2.$$ This is
impossible since $H(t)$ is a decreasing function  and $H(t_0)\geq
\eta.$ In the same way, $\lim_{t\rightarrow\infty}\phi=-\infty$
leads to a contradiction. Then, $\lim_{t\rightarrow\infty}\phi=0$
and this implies $\lim_{t\rightarrow\infty}V(\phi(t))=0,$  and
again by \eqref{newFried}, $\lim_{t\rightarrow\infty}H(t)=0.$

Thus, we have proved that if the potential has a local minimum at zero, if the derivative of
the potential is bounded in the same set where the potential
itself is, and provided the derivative of the logarithm of the
coupling function is bounded by above, then, the energy densities of
DM and radiation, and the kinetic energy density of DE tend to zero as the time
goes forward.  Hence, the Universe would expand forever in a de Sitter phase.
Also we have proved, in a similar way as in Proposition 3 in
\cite{Miritzis:2003ym}, that under the additional assumption of
$V(\phi)$ being strictly decreasing (increasing) if $\phi<0$
($\phi>0$), then the scalar field can be either zero
or divergent into the future (the former case holds if the Hubble
scalar vanishes asymptotically).

In order to complement the former ideas, let us consider a non-negative potential with no necessarily a local
minimum at $(0,0)$, and let's us find conditions for the stability of de Sitter solutions, and  let's characterize the asymptotic properties of the scalar field at late times.

\begin{prop}\label{Proposition III}
Suppose that there exists a nonzero constant $K,$ such that
$\chi'(\phi)/\chi(\phi)\leq 2 K/(2-\gamma)(4-3\gamma).$ Let  $V$
be a potential function with the properties:
 \begin{enumerate}
    \item $V\geq 0$ and $\lim_{\phi\rightarrow -\infty} V(\phi)=+\infty.$
    \item $V'$ is continuous and $V'(\phi)<0.$
    \item If  $A\subset\mathbb{R}$ is such that $V$ is bounded in $A,$ Then, $ V'(\phi)$ is bounded in $A.$
 \end{enumerate}
Then, $\lim_{t\rightarrow\infty} (\rho_m,\rho_r,y)=(0,0,0),$ and
$\lim_{t\rightarrow\infty}\phi=+\infty.$
\end{prop}

{\bf Proof}. From equations \eqref{newconsm} and \eqref{newconsr}, follow that the sets $\rho_m>0$ and $\rho_r>0$ are
invariant under the flow of \eqref{Syst:13} with
restriction \eqref{newFried}; besides $\rho_m$ and $\rho_r$ are different from zero if
$\rho_m(t_0)$ and $\rho_r(t_0)$ are they are at the initial time. From this
fact we have that $H$ is never zero (and thus do not have changes of sign) since
by \eqref{newFried}, $3 H(t)^2\geq \rho_m(t)>0$ for all $t>t_0,$ then,
$H$ is always nonnegative if initially it is. Besides, from
equation \eqref{newRaych}, follows that $H$ is decreasing, then
$\exists \lim_{t\rightarrow\infty} H(t)=\eta\geq 0$ and
$$\frac{1}{2}\int_{t_0}^{\infty}\left(y^2+\gamma\rho_m+\frac{4}{3}\rho_r\right)
d t = H(t_0)-\eta<+\infty.$$ As in proposition \ref{Proposition
I}, the total time derivative of $y^2+\gamma\rho_m +\frac{4}{3}\rho_r$ is bounded.
Hence $\lim_{t\rightarrow\infty}(\rho_m, \rho_r, y)=(0,0,0).$

It can be proved that $\lim_{t\rightarrow\infty}\phi=+\infty$ in
the same way as proved in \ref{Proposition II}.

From equation \eqref{newFried} we have that $\lim_{t\rightarrow\infty}
V(\phi)=3\eta^2.$ Since $V$ is strictly decreasing with respect to
$\phi;$ then $V(\phi)>\lim_{\phi\rightarrow\infty} V(\phi)$ for
all $\phi,$ therefore $\lim_{t\rightarrow\infty} V(\phi(t))\geq
\lim_{\phi\rightarrow\infty} V(\phi).$ We will consider two cases:
 \begin{enumerate}
   \item If $\lim_{t\rightarrow\infty} V(\phi(t))= \lim_{\phi\rightarrow\infty} V(\phi),$
   by the continuity of  $V$ is obvious that
$\lim_{t\rightarrow\infty}\phi=+\infty;$
   \item If  $ \lim_{t\rightarrow\infty} V(\phi(t))> \lim_{\phi\rightarrow\infty} V(\phi),$
   then, there exists a unique $\bar{\phi}$ such that
   $$\lim_{t\rightarrow\infty} V(\phi(t))=V(\bar{\phi}).$$ Since  $V$ is continuous and
   strictly decreasing follows that $$\lim_{t\rightarrow\infty} \phi=\bar{\phi}.$$
From equation \eqref{newKG} follows that
$$\lim_{t\rightarrow\infty}\frac{d}{d
t}y=-V'(\bar{\phi})>0,$$ therefore, exists $t'$ such that
$\frac{d}{d t}y>-V'(\bar{\phi})/2$ for all $ t\geq t'.$ From this
fact we conclude that
$$y(t)- y(t')
>-\frac{V'(\bar{\phi})}{2}(t- t'),$$ which is impossible since
$\lim_{t\rightarrow\infty}y(t)=0.$ Finally
$\lim_{t\rightarrow\infty}\phi=+\infty.$ $\blacksquare$
 \end{enumerate} 

If additionally, the potential is such that
$\lim_{\phi\rightarrow\infty} V(\phi)=0,$ then we conclude that
$H\rightarrow 0$ as $t \rightarrow\infty.$

The previous results are extensions of the Remark 1 and Propositions 4, 5 and 6 discussed in \cite{Lap_Lambert} when the radiation is included in the cosmic budget.

For completeness let's show our Proposition 3 in \cite{Nova_Science}, which is an extension of the Proposition 1 of
\cite{Giambo':2009cc} for flat FLRW models since we have included radiation. This proposition gives a characterization of the
future attractor of the system \eqref{Syst:13} under some mild assumptions for the potential.

First, let us formalize notion of degenerate local minimum introduced in
\cite{Giambo':2009cc}:

\begin{defn}
The function $V(\phi)$ is said to have a degenerate local minimum
at $\phi_\star$ if  $$V'(\phi),V''(\phi), \dots V^{(2n-1)}$$
vanish at $\phi_*,$ and $V^{(2n)}(\phi_*)>0,$ for some integer
$n.$
\end{defn}

Then, we have the proposition: 
\begin{prop}[Proposition 3 in \cite{Nova_Science}]\label{thmIII} Suppose that $V(\phi)\in C^2(\mathbb{R})$ satisfies the following conditions \footnote{Empty set is bounded and finite, and it is not excluded in the hypothesis (i) and (ii).}:
\begin{itemize}
\item[(i)] The set $\{\phi: V(\phi)<0\}$ is
bounded; \item[(ii)] The set of singular points
of $V(\phi)$ is finite.
\end{itemize}
Let $\phi_*$ a strict local minimum (possibly
degenerate) for $V(\phi),$ with $V(\phi_*)\geq 0$. Then ${\bf
p}_*:=\left(\phi_*,y_*=0,{\rho_m}_*=0,\rho_r=0,
H=\sqrt{\frac{V(\phi_*)}{3}}\right)$ is an asymptotically stable
singular point for the flow of \eqref{Syst:13}.
\end{prop}

{\bf Proof.}

The demonstration proceeds in an analogous way as the proof of the Proposition 1 in
\cite{Giambo':2009cc}. The main difference is that we have considered radiation but flat FLRW geometry ($k=0$). In this case the function 
\be
W(\phi,y,\rho_m,H)\equiv H^2-\frac{1}{3}\left(\frac{1}{2}y^2+V(\phi)+\rho_m\right)=\frac{1}{3}\rho_r
\ee evolves like 
\be\dot W=-4 H W\ee which decays more faster to zero than the function $W(\phi,y,\rho_m,H)=-k a^{-2},\, k=-1,0$ defined in \cite{Giambo':2009cc} as $a\rightarrow +\infty$. (The complete proof is offered in \cite{Nova_Science}).
$\blacksquare$

\section{Dynamical analysis for $\phi\rightarrow +\infty$.}\label{Sect:Inf}

In this section we will investigate the flow as
$\phi\rightarrow \infty$ following the nomenclature and formalism introduced in \cite{Foster:1998sk} (see also \cite{Giambo:2008sa} and \cite{Nova_Science}). Analogous results hold as $\phi\rightarrow-\infty.$

\begin{defn}[Function well-behaved at infinity \cite{Foster:1998sk}]\label{WBI}
Let $V:\mathbb{R}\rightarrow \mathbb{R}$ be a $C^2$ non-negative
function. Let there exist some $\phi_0>0$ for which $V(\phi)> 0$
 for all $\phi>\phi_0$ and some number $N$ such that the function
$W_V:[\phi_0,\infty)\rightarrow \mathbb{R}$,
$$ W_V(\phi)=\frac{V'(\phi)}{V(\phi)} - N $$
 satisfies
\begin{equation}
\lim_{\phi\rightarrow\infty}W_V(\phi)=0.\label{Lim}
\end{equation}
Then we say that $V$ is Well Behaved at Infinity (WBI) of
exponential order $N$.
\end{defn}

\begin{thm}[Theorem 2, \cite{Foster:1998sk}]\label{thm2.6}
Let $V$ be a WBI function of exponential order $N,$ then, for all $\lambda>N,$
$$\lim_{\phi\rightarrow +\infty}e^{-\lambda \phi}V(\phi)=0.$$
\end{thm}

In order to classify the smoothness of WBI functions at infinity
it is introduced the definition

\begin{defn}\label{bar}
Let be some coordinate transformation $\varphi=f(\phi)$ mapping a
neighborhood of infinity to a neighborhood of the origin. If $g$
is a function of $\phi$,
 $\overline{g}$ is the function of $\varphi$ whose domain is the range of
$f$ plus the origin, which takes the values;

$$
\overline{g}(\varphi)=\left\{\begin{array}{rcr} g(f^{-1}(\varphi))&,&\varphi>0\\
                                     \lim_{\phi\rightarrow\infty} g(\phi)&,&\varphi=0 \end{array}\right.
$$
\end{defn}
\begin{defn}[Class k WBI functions \cite{Foster:1998sk}]\label{CkWBI}
A $C^k$ function $V$ is class k WBI if it is WBI and if there
exists $\phi_0>0$ and a coordinate transformation
$\varphi=f(\phi)$ which maps the interval $[\phi_0,\infty)$ onto
$(0, \epsilon]$, where $\epsilon=f(\phi_0)$ and
$\lim_{\phi\rightarrow\infty} f=0$, with  the following additional
properties:
\begin{tabbing}
i)\hspace{0.4cm}\=  $f$ is $C^{k+1}$ and strictly decreasing.\\
ii)            \>the functions $\overline{W_V}(\varphi)$ and
$\overline{f'}(\varphi)$ are $C^k$ on
the  closed interval $[0,\epsilon]$.\\
iii)           \> ${\displaystyle
\frac{\mathrm{d}\overline{W_V}}{\mathrm{d}\varphi}(0)=\frac{\mathrm{d}\overline{f'}}{\mathrm{d}\varphi}(0)=0.}$
\end{tabbing}
\end{defn}

We designate the set of all class k WBI functions ${\cal E}^k_+.$

By assuming that $V,\chi\in {\cal E}^3_+,$ with exponential orders
$N$ and $M$ respectively, we can define a dynamical system well
suited to investigate the dynamics near the initial singularity.
We will investigate the singular points therein. Particularly
those representing scaling solutions and those associated with the
initial singularity \cite{Nova_Science}.

Let's define the new Hubble-normalized dimensionless variables \cite{Nova_Science}:
\be \sigma_1=\phi,
\sigma_2=\frac{\dot\phi}{\sqrt{6}H},\,\sigma_3=\frac{\sqrt{\rho_m}}{\sqrt{3}H},\,\sigma_4=\frac{\sqrt{V}}{\sqrt{3}H},
\,\sigma_5=\frac{\sqrt{\rho_r}}{\sqrt{3}H}\label{vars}\ee and the time coordinate \be \mathrm{d}\tau=3 H \mathrm{d}t.\ee

Using these coordinates the equations \eqref{Syst:13}
recast as an autonomous system satisfying an inequality arising
from the Friedmann equation \eqref{Fried}. This system is given by \cite{Nova_Science}:
\begin{subequations}\label{Syst:22}
\begin{align}
&\sigma_1'=\sqrt{\frac{2}{3}} \sigma_2 \label{eq0phi}\\
&\sigma_2'=\sigma_2^3+\frac{1}{6}\left(3\gamma \sigma_3^2+4 \sigma_5^2-6\right)\sigma_2
-\frac{\sigma_4^2}{\sqrt{6}}\frac{\mathrm{d}\ln V(\sigma_1)}{\mathrm{d}\sigma_1}
+\frac{\left(4-3\gamma\right)\sigma_3^2}{2\sqrt{6}}\frac{\mathrm{d}\ln \chi(\sigma_1)}{\mathrm{d}\sigma_1},\label{eq0x1}\\
&\sigma_3'=\frac{1}{6}\sigma_3\left(6\sigma_2^2+3\gamma\left(\sigma_3^2-1\right)+4 \sigma_5^2\right)
-\frac{\left(4-3\gamma\right)\sigma_2 \sigma_3}{2\sqrt{6}}\frac{\mathrm{d}\ln \chi(\sigma_1)}{\mathrm{d}\sigma_1},\label{eq0x2}\\
&\sigma_4'=\frac{1}{6}\sigma_4\left(6\sigma_2^2+3\gamma
\sigma_3^2+4 \sigma_5^2\right)
+\frac{\sqrt{6}}{6}\sigma_2 \sigma_4 \frac{\mathrm{d}\ln V(\sigma_1)}{\mathrm{d}\sigma_1},\label{eq0x3}\\
&\sigma_5'=\frac{1}{6}\sigma_5\left(6\sigma_2^2+3\gamma
\sigma_3^2+4 \sigma_5^2-4\right).\label{eq0x4}
\end{align}
\end{subequations}
The system \eqref{Syst:22} defines a flow 
in the phase space
\be \Sigma :=\left\{\sigma\in\mathbb{R}^5: \sum_{j=2}^5
\sigma_j^2=1, \sigma_j\geq 0, j=3, 4, 5\right\}\label{Sigma}.\ee

Let $\Sigma_\epsilon=\left\{(\sigma_1,
\sigma_2,\sigma_3,\sigma_4,\sigma_5)\in \Sigma:
\sigma_1>\epsilon^{-1}\right\}$ where $\epsilon$ is any positive constant which is chosen sufficiently small so as to avoid any points where $V$ or $\chi=0,$ thereby ensuring that $\overline{W_V}(\varphi)$ and $\overline{W}_{\chi}(\varphi)$ are well-defined. \footnote{See \ref{bar} for the definition of functions with bar.}

Let be defined the projection map
\ben && \pi_1: \Sigma_\epsilon \rightarrow \Omega_\epsilon \nonumber\\
&&(\sigma_1, \sigma_2,\sigma_3,\sigma_4,\sigma_5)\rightarrow
(\sigma_1, \sigma_2,\sigma_4,\sigma_5)\een where
\be\Omega_\epsilon:=\left\{\sigma\in\mathbb{R}^4:\sigma_1>\epsilon^{-1},
\sigma_2^2+\sigma_4^2+\sigma_5^2\leq 1, \sigma_j\geq 0, j= 4,
5\right\}.\ee

Let be defined in $\Omega_\epsilon$ the coordinate transformation
$(\sigma_1, \sigma_2,\sigma_4,\sigma_5)
\stackrel{\varphi=f(\sigma_1)}{\longrightarrow} (\varphi,
\sigma_2,\sigma_4,\sigma_5)$ where $f(\sigma_1)$ tends to zero as
$\sigma_1$ tends to $+\infty$ and has been chosen so that the conditions i)-iii) of definition \ref{CkWBI} are satisfied with
$k=2.$

The flow of \eqref{Syst:22} defined on
$\Sigma_\epsilon$ is topologically equivalent (under $f\circ
\pi_1$) to the flow of the 4-dimensional dynamical system \cite{Nova_Science}:
\begin{subequations}\label{Syst:26}
\begin{align}
&\varphi'=\sqrt{\frac{2}{3}} \overline{f'} \sigma_2,\label{eqvphi}\\
&\sigma_2'= \sigma_2^3+\left(\frac{2 \sigma_5^2}{3}-1\right)
\sigma_2-\frac{\left(\overline{W_V}+N\right)
\sigma_4^2}{\sqrt{6}}+\nonumber\\&+\left(\frac{\left(\overline{W}_{\chi}+M\right)
(4-3 \gamma )}{2 \sqrt{6}}+\frac{\sigma_2 \gamma }{2}\right)
   \left(1-\sigma_2^2-\sigma_4^2-\sigma_5^2\right)\label{eqxrad}\\
&\sigma_4'=\frac{1}{6} \sigma_4 \left(\sqrt{6} \left(\overline{W_V}+N\right) \sigma_2+3(2-\gamma)\sigma_2^2+3\gamma(1-\sigma_4^2)+(4-3\gamma)\sigma_5^2\right), \label{eqyrad}\\
&\sigma_5'=\frac{1}{6} \sigma_5 \left(3(2-\gamma)\sigma_2^2-3\gamma \sigma_4^2-(4-3\gamma)(1-\sigma_5^2)\right),
\label{eqzrad}
\end{align}
\end{subequations}
defined in the phase space \footnote{For notational simplicity we will denote the image of $\Omega_\epsilon$ under $f$ by the same symbol.} \be\Omega_\epsilon=\{(\varphi, \sigma_2,
\sigma_4,\sigma_5)\in\mathbb{R}^4: 0\leq\varphi\leq
f(\epsilon^{-1}), \sigma_2^2+\sigma_4^2+\sigma_5^2\leq 1,
\sigma_4\geq 0, \sigma_5\geq 0\}.\label{PhaseSpace}\ee It can be easily proved that \eqref{PhaseSpace} defines a manifold with boundary of dimension 4. Its boundary, $\partial\Omega_\epsilon,$  is the union of the sets $\{p\in\Omega_\epsilon:
\varphi=0\},\,\{p\in\Omega_\epsilon:
\varphi=f(\epsilon^{-1})\},\,\{p\in\Omega_\epsilon:
\sigma_4=0\},\,\{p\in\Omega_\epsilon: \sigma_5=0\}$ with the
unitary 3-sphere.
The coordinates, existence conditions and stability of the singular points of the flow of \eqref{Syst:26} in the phase space \eqref{PhaseSpace} are presented in the  \ref{AppendixA}. Finally, the physical description of the solutions and connection with observables is discussed in \ref{AppendixB}. The results discussed in both appendices were first published in our reference \cite{Nova_Science}. 
\section{Examples}\label{applications}

In this section we discuss the example of a double exponential potential, presented in \cite{Tzanni:2014eja}, applying the procedure of \cite{Foster:1998sk,Nova_Science}. Our new results complement those in \cite{Tzanni:2014eja}. Next we revisit the example of a powerlaw coupling function and an Albrecht-Skordis potential, first introduced in \cite{Leon:2008de}, and then extended in section 4.4 of \cite{Nova_Science}.

\subsection{Double exponential potential and exponential coupling function}\label{Sect:double_exp}

First let's discuss the case presented in \cite{Tzanni:2014eja}. 
In this example the potential is double exponential: $V(\phi)=V_1 e^{-\alpha \phi }+V_2 e^{-\beta\phi}, \; 0<\alpha<\beta,$ and $\chi=\chi_0 \exp\left[\frac{\lambda\phi}{4-3\gamma}\right]$, where $\lambda$ is a constant. Then, for $\gamma<\frac{4}{3}$, choose $K\geq \frac{(2-\gamma)\lambda}{2}$, and for $\frac{4}{3}<\gamma\leq 2$, choose $K\leq \frac{(2-\gamma)\lambda}{2}.$ Thus, the hypothesis for $\chi$ in our results it is easily fulfilled. Since $V_1>0, V_2>0,$ we have $V'(\phi)<0<V(\phi), \forall \phi.$ Furthermore, under the hypothesis $0<\alpha<\beta,$ we have the strong restriction $-\beta V(\phi)<V'(\phi)<-\alpha V(\phi), \forall \phi.$ Thus, the hypothesis of our Proposition \ref{Proposition III} are satisfied and we result in the Proposition 1 of \cite{Tzanni:2014eja}.

Additionally we have that under the hypothesis  $0<\alpha<\beta,$ $V(\phi)=V_1 e^{-\alpha \phi }+V_2 e^{-\beta\phi}$ is well-behaved at $\phi\rightarrow +\infty$ with exponential order $N=-\alpha$. On the other hand the coupling function is well behaved of exponential order $M=\frac{\lambda}{4-3\gamma}.$ 
Under the transformation $\varphi=\phi^{-1}$ we have that $V,\chi\in {\cal E}^k_+,$ and 
\begin{align}
&\overline{W_\chi}(\vphi)=0, \label{WchiAS1}\\
&\overline{W_V}(\vphi)= \left\{\begin{array}{rcr} \frac{V_2 (\alpha -\beta ) e^{\frac{\alpha }{\vphi}}}{V_1
   e^{\frac{\beta }{\vphi}}+V_2 e^{\frac{\alpha }{\varphi}}}&,&\vphi>0\\
                                     0 &,&\vphi=0 \end{array}\right.\label{WVAS1},\\ 
&\overline{f'}(\vphi)=-\vphi^2,\label{fAS1}																
\end{align}

In this example, the evolution equations for $\varphi,$
$\sigma_2,$ $\sigma_4,$ and $\sigma_5$ are given by the equations
\eqref{Syst:26} with $M=\frac{\lambda}{4-3\gamma},\,N=-\alpha,$ and
$\overline{W}_{\chi}(\varphi),\,\overline{W}_{V}(\varphi),$ and
$\overline{f'},$ given by \eqref{WchiAS1}, \eqref{WVAS1} and
\eqref{fAS1} respectively. The state space is defined by
$$\Omega_\epsilon=\{(\varphi, \sigma_2,
\sigma_4,\sigma_5)\in\mathbb{R}^4: 0\leq\varphi\leq
{\epsilon}, \sigma_2^2+\sigma_4^2+\sigma_5^2\leq 1,
\sigma_4\geq 0, \sigma_5\geq 0\}.$$

Now, since the function $\overline{W_V}$ defined by \eqref{WVAS1}  is a transcendental function for $\vphi=0,$ we can introduce the new variable
\be v=-\frac{V_2 (\alpha -\beta ) e^{\frac{\alpha }{\vphi}}}{V_1
   e^{\frac{\beta }{\vphi}}+V_2 e^{\frac{\alpha }{\varphi}}}\geq 0  \label{NewV}
	\ee which is better for numerical integrations. It can be proved that under the hypothesis $0<\alpha<\beta,$ $v\rightarrow 0^+$ as $\vphi\rightarrow 0^+.$

Using the coordinate transformation \eqref{NewV} we obtain the dynamical system
\begin{subequations}\label{Example1}
\begin{align}
&v'=\sqrt{\frac{2}{3}} \sigma _2 v (\alpha -\beta +v),\\
&\sigma_2'= \frac{\alpha  \sigma _4^2}{\sqrt{6}}+\frac{1}{6} \sigma _2 \left(-3 \gamma  \left(\sigma _2^2+\sigma _4^2+\sigma _5^2-1\right)+6 \sigma _2^2+4
   \sigma _5^2-6\right)-\frac{\lambda  \left(\sigma _2^2+\sigma _4^2+\sigma _5^2-1\right)}{2 \sqrt{6}},\\
&\sigma_4'=-\frac{1}{6} \sigma _4 \left(\sqrt{6} \alpha 
   \sigma _2+3 (\gamma -2) \sigma _2^2+3 \gamma  \left(\sigma _4^2+\sigma _5^2-1\right)-4 \sigma _5^2\right)-\frac{\sigma _2 \sigma _4
   v}{\sqrt{6}},\\
&\sigma_5'=\frac{1}{6} \sigma _5 \left(-3 \gamma  \left(\sigma _2^2+\sigma _4^2+\sigma _5^2-1\right)+6 \sigma _2^2+4 \sigma
   _5^2-4\right),	
\end{align}
\end{subequations}
defined in the 
phase space 
$$\Psi_\epsilon=\{(v, \sigma_2,
\sigma_4,\sigma_5)\in\mathbb{R}^4: 0\leq v\leq
\frac{V_2 (\beta -\alpha ) e^{\alpha /\epsilon }}{V_1 e^{\beta /\epsilon }+V_2 e^{\alpha /\epsilon }}, \sigma_2^2+\sigma_4^2+\sigma_5^2\leq 1,
\sigma_4\geq 0, \sigma_5\geq 0\}.$$

\begin{table}[t]
\caption{\label{crit} Location of the singular points of the flow
of \eqref{Example1} defined in the invariant set
$\left\{p\in\Omega_\epsilon: \varphi=0\right\}$ for $M=\frac{\lambda}{4-3\gamma}$ and
$N=-\alpha.$ We use the definitions $\Gamma^+(\alpha,\gamma)=\alpha\pm \sqrt{\alpha^2+6 \gamma ^2-12 \gamma },$ and $\Upsilon(\gamma)=\sqrt{2} \sqrt{(4-3 \gamma ) (2-\gamma )}.$}\bigskip 
\resizebox{1.0\textwidth}{!}
{\begin{tabular}[t]{|l|c|c|c|c|}
\hline
Label&$(\sigma_2,\sigma_4,\sigma_5)$&Existence&Stability$^{\rm a}$\\[2ex]
\hline
$P_1$&$(-1,0,0)$&always & unstable if $\alpha >-\sqrt{6}, \lambda >-\sqrt{6} (2-\gamma)$\\[2ex]
&&& saddle otherwise \\
\hline
$P_2$&$(1,0,0)$&always & unstable if $\alpha <\sqrt{6}, \lambda <\sqrt{6} (2-\gamma)$\\[2ex]
&&& saddle otherwise \\
\hline
$P_3$&$\left(\frac{\lambda}{\sqrt{6}(2-\gamma)},0,0\right)$&$\lambda^2\leq 6(2-\gamma)^2$ & stable for \\[2ex]
&&& $\left\{\begin{array}{c}  0<\gamma \leq \frac{2}{3},\, \text{and}\\
\sqrt{6(2-\gamma ) \gamma }<\alpha \leq \frac{\sqrt{8(2-\gamma)}}{\sqrt{4-3 \gamma}},\, \text{and}\\ 
\Gamma^-(\alpha,\gamma)<\lambda <\Gamma^+(\alpha,\gamma) \end{array}\right.$  \\[2ex]
	&&& or $\left\{\begin{array}{c} 0<\gamma \leq \frac{2}{3},\, \text{and}	\\
	\alpha >\frac{\sqrt{8(2-\gamma)}}{\sqrt{4-3 \gamma}},\, \text{and}	\\ 
	\Gamma ^-(\alpha ,\gamma )<\lambda <\Upsilon (\gamma )\end{array}\right.$ \\[2ex]
	&&& or $\left\{\begin{array}{c}\frac{2}{3}<\gamma <\frac{4}{3},\, \text{and}\\
	\alpha >\frac{\sqrt{8(2-\gamma)}}{\sqrt{4-3 \gamma}},\, \text{and}\\
	\Gamma ^-(\alpha ,\gamma )<\lambda <\Upsilon (\gamma )\end{array} \right.$\\[2ex]
	&&& saddle otherwise \\[2ex]
\hline
$R_1$&$(0,0,1)$&always & saddle\\[2ex]
\hline
$P_4$&$\left(\frac{\alpha}{\sqrt{6}},\sqrt{1-\frac{\alpha^2}{6}},0\right)$&$\alpha^2<6$ & stable for $0<\alpha <2, \lambda >\frac{2 \alpha ^2-6 \gamma }{\alpha }$\\
&&& saddle otherwise\\[2ex]
\hline
$P_{5,6}$&$\left(\frac{\sqrt{6} \gamma }{2 \alpha -\lambda },\pm \frac{\sqrt{-2 \alpha  \lambda -6 (\gamma -2) \gamma +\lambda ^2}}{\lambda -2 \alpha
   },0\right)$ & $\frac{\alpha  (\lambda -2 \alpha )+6 \gamma }{(\lambda -2 \alpha )^2}\leq 0$ & Numerical inspection\\[2ex]
\hline
$R_2$ & $\left( \frac{\sqrt{\frac{2}{3}} (4-3 \gamma )}{\lambda },0,\frac{\sqrt{-6 \gamma ^2+20 \gamma +\lambda ^2-16}}{|\lambda| }\right)$ &  $\gamma <\frac{4}{3},\lambda ^2\geq \Upsilon(\gamma)^2$ & stable for \\
&&& 
$\left\{\begin{array}{c}\gamma<\frac{4}{3},\\
 \alpha>\sqrt{\frac{8(2-\gamma)}{4-3\gamma}},\\
 \Upsilon(\gamma)<\lambda< \frac{1}{2} \alpha  (4-3 \gamma)
	\end{array}\right.$\\
	&&& saddle otherwise\\[2ex]\hline
$R_3$ & $\left(\frac{2\sqrt{\frac{2}{3}}}{\alpha},\frac{2}{\sqrt{3}\alpha},\frac{\sqrt{\alpha^2-4}}{\alpha}\right)$& $\alpha\geq 2$ & stable if $\alpha >2,\lambda >\frac{1}{2} \alpha  (4-3 \gamma),$ \\
&&& saddle otherwise \\[2ex]
\hline
\end{tabular}}
$^{\rm a}$ The stability is analyzed for the flow restricted to
the invariant set $v=0$.
\end{table}

In
table \ref{crit} are summarized the location, existence conditions
and stability of the singular points of the system \eqref{Example1}. The stability is
analyzed for the flow restricted to the invariant set $v=0$,
i.e., we are not taking into account perturbations along the
$v$-axis.

Let us discuss some physical properties of the cosmological
solutions associated to the singular points displayed in table
\ref{crit}.
\begin{itemize}
\item  $P_{1,2}$  represent kinetic-dominated cosmological
solutions. They behave as stiff-like matter. The associated
cosmological solution satisfies $H=\frac{1}{3 t-c_1},a=\sqrt[3]{3
t-c_1} c_2,\phi =c_3\pm\sqrt{\frac{2}{3}} \ln \left(3
   t-c_1\right),$ where $c_j,\,j=1,2,3$ are integration constants. These solutions are associated with the local past attractors of the system for an open set of values of the parameters $\alpha$ and $\lambda$.
\item $P_3$	represents a matter-kinetic scaling solution. The associated asymptotic cosmological solutions satisfy the rates $H=\frac{8-4 \gamma }{4 (\gamma -2) c_1+t \left(\lambda ^2-6 (\gamma -2) \gamma \right)},a=c_2 \left(4 (\gamma -2) c_1+t \left(\lambda ^2-6
   (\gamma -2) \gamma \right)\right){}^{-\frac{4 (\gamma -2)}{\lambda ^2-6 (\gamma -2) \gamma }},\rho_m=\frac{48 (\gamma -2)^2-8 \lambda ^2}{\left(t
   \left(6 (\gamma -2) \gamma -\lambda ^2\right)-4 (\gamma -2) c_1\right){}^2}+c_3,\phi=\ln \left[\left(4 (\gamma -2) c_1+t \left(\lambda
   ^2-6 (\gamma -2) \gamma \right)\right)^\frac{4 \lambda }{\lambda ^2-6 (\gamma -2) \gamma }\right]+c_4$, where $c_j,\,j=1,2,3,4$ are integration constants. These solutions are stable or saddle depending on the parameters $\alpha$ and $\gamma$. 
	\item Point $R_1$ represents a radiation dominated solution, ans asymptotically the cosmological solutions satisfy the rates
	$H=\frac{1}{2 t-c_1},a=c_2 \sqrt{2 t-c_1},\rho_r=c_3-\frac{6}{c_1-2 t}.$ It is always a saddle point. 
	
	\item $P_4$ represents power-law scalar-field dominated inflationary
cosmological solutions. As $\phi\rightarrow +\infty$ the potential
behaves as $V\sim V_1 \exp[-\alpha \phi]$. Thus it is easy to obtain
the asymptotic exact solution: $H=\frac{2}{\alpha ^2 t-2 c_1},a=c_2 \left(\alpha ^2 t-2 c_1\right){}^{\frac{2}{\alpha ^2}},\phi=\ln \left[\left(\alpha ^2 t-2
   c_1\right)^\frac{2}{\alpha }\right]+c_3$.
	
	\item $P_{5,6}$ represent matter-kinetic-potential scaling solutions. In the limit $\phi\rightarrow +\infty$ 
we have the asymptotic expansions: $H=\frac{2 \alpha -\lambda }{-2 \alpha  c_1+c_1 \lambda +3 \alpha  \gamma  t},a=c_2 \left(c_1 (\lambda -2 \alpha )+3 \alpha  \gamma 
   t\right){}^{\frac{2 \alpha -\lambda }{3 \alpha  \gamma }},\rho_m =\frac{6 \left(2 \alpha ^2-\alpha  \lambda -6 \gamma \right)}{\left(c_1 (\lambda -2
   \alpha )+3 \alpha  \gamma  t\right){}^2}+c_3,\phi =\ln \left[\left(c_1 (\lambda -2 \alpha )+3 \alpha  \gamma  t\right)^\frac{2}{\alpha }\right]+c_4.$
	
	\item $R_3$ represents a radiation-matter-scalar field scaling solution. In the limit $\phi\rightarrow +\infty$ 
we have the asymptotic expansions: $H=\frac{1}{2 t-c_1},a=c_2 \sqrt{2 t-c_1},\rho_m =\frac{16-12 \gamma }{\lambda ^2 \left(c_1-2 t\right){}^2}+c_3,\rho_r=c_4-\frac{6 \left(-6
   \gamma ^2+20 \gamma +\lambda ^2-16\right)}{\lambda ^2 \left(c_1-2 t\right)},\phi =\ln\left[\left(\lambda  \left(2
   t-c_1\right)\right)^\frac{(4-3 \gamma )}{\lambda}\right]+c_5.$
	
	\item $R_3$ represents radiation-kinetic-potential scaling solutions. The associated cosmological solutions satisfy the asymptotic expansions: $H=\frac{1}{2 t-c_1},a=c_2 \sqrt{2 t-c_1},\rho_r=c_3-\frac{6 \left(\alpha ^2-4\right)}{\alpha ^2 \left(c_1-2 t\right)},\phi=
   \ln \left[\left(2 \alpha  t-\alpha  c_1\right)^\frac{2}{\alpha }\right]+c_4$ as $\phi\rightarrow +\infty.$
\end{itemize}

In order to illustrate our analytical results we proceed to some numerical experimentation. 

\begin{figure}[t]
\begin{center}
\hspace{0.8cm}
\includegraphics[scale=0.4]{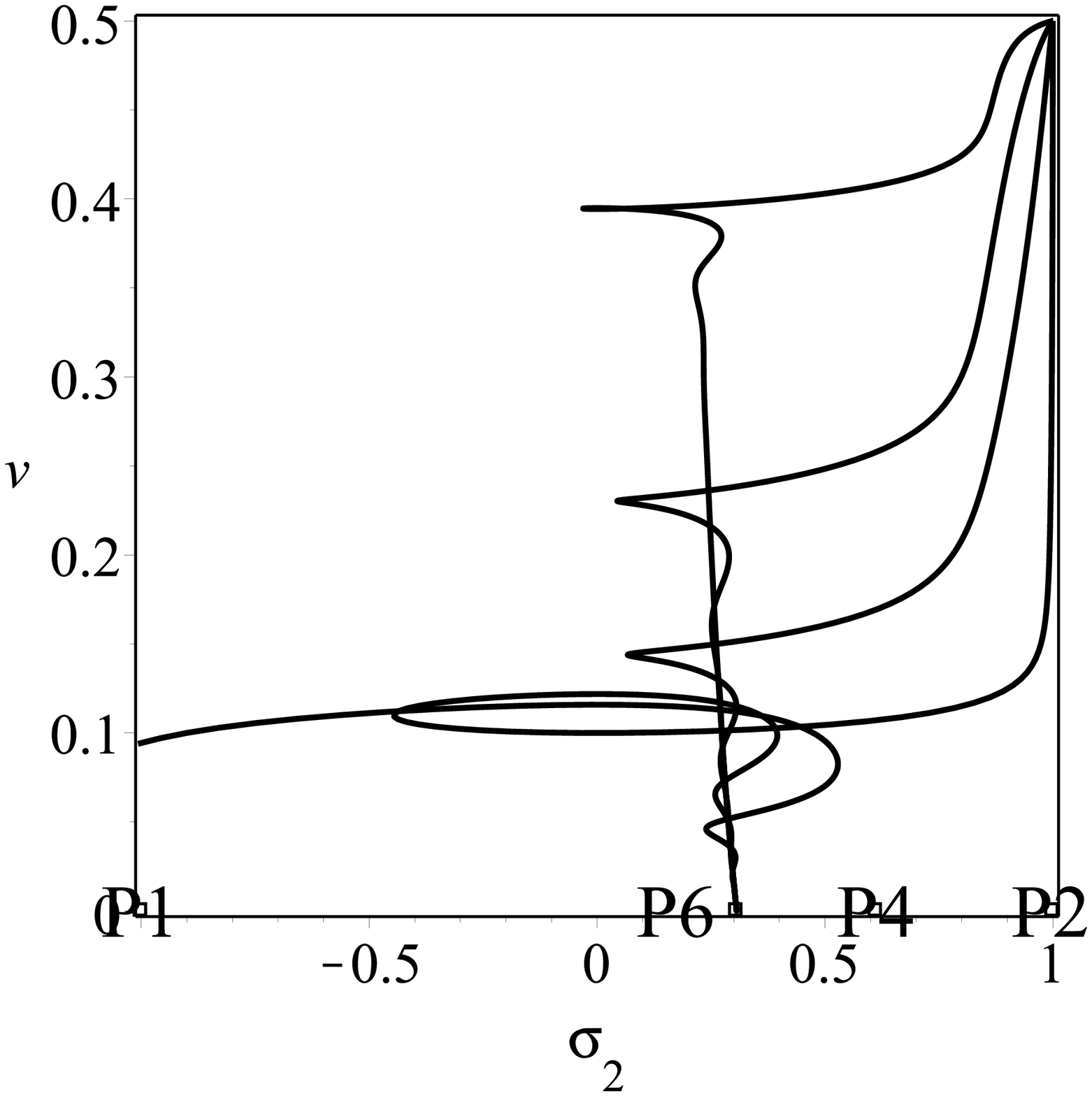}
\caption[Projection in the plane $(\sigma_2, v)$ of the flow of \eqref{Example1} restricted to the invariant set $\{\sigma_5=0\}\subset \bar{\Psi}_\epsilon$.]{Projection in the plane $(\sigma_2, v)$ of the flow of \eqref{Example1} restricted to the invariant set $\{\sigma_5=0\}\subset \bar{\Psi}_\epsilon$ for the potential  $V(\phi)=V_1 e^{-\alpha \phi }+V_2 e^{-\beta\phi},$ and the coupling function $\chi=\chi_0 \exp\left[\frac{\lambda\phi}{4-3\gamma}\right]$.  We select the values of the parameters: $V_1=1, V_2=2, \alpha=1.5, \beta=2, \lambda=-5$ and $\gamma=1.$ In the figure it is illustrated the stability of  $P_6$ along the $v$-axis.} \label{FIG1}
\end{center}
\end{figure}

\begin{figure}[t]
\begin{center}
\hspace{0.8cm}
\includegraphics[scale=0.4]{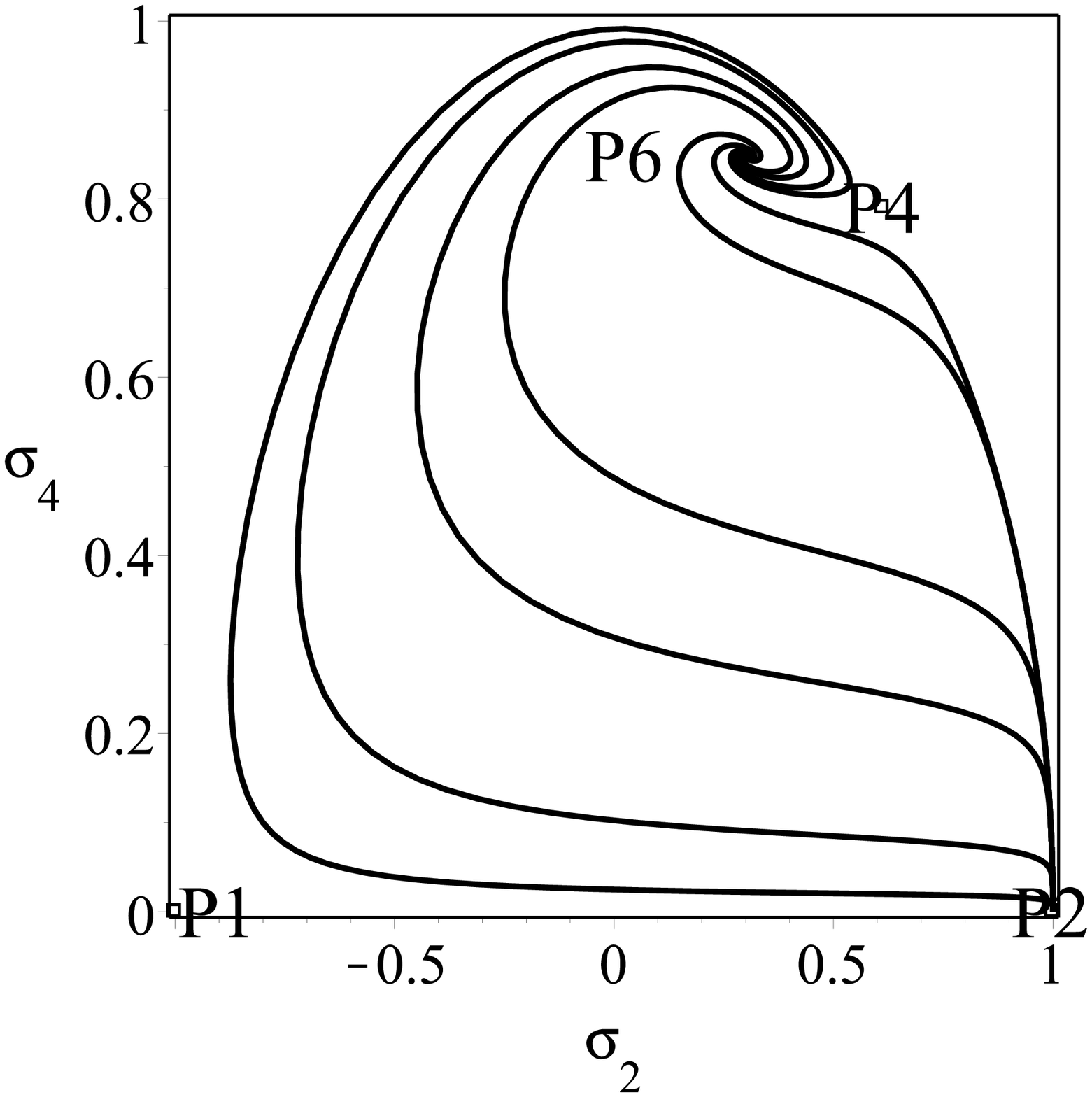}
\caption[Projection in the plane $(\sigma_2, \sigma_4)$ of the flow of \eqref{Example1} restricted to the invariant set $\{\sigma_5=0\}\subset \bar{\Psi}_\epsilon$.]{Projection in the plane $(\sigma_2, \sigma_4)$ of the flow of \eqref{Example1} restricted to the invariant set $\{\sigma_5=0\}\subset \bar{\Psi}_\epsilon$ for the potential  $V(\phi)=V_1 e^{-\alpha \phi }+V_2 e^{-\beta\phi},$ and the coupling function $\chi=\chi_0 \exp\left[\frac{\lambda\phi}{4-3\gamma}\right]$.  We select the values of the parameters: $V_1=1, V_2=2, \alpha=1.5, \beta=2, \lambda=-5$ and $\gamma=1.$ In the figure it is illustrated the stability of $P_6$. $P_1$ is a source and $P_2$ and $P_4$ are saddles. $P_3$ does not exists.} \label{FIG2}
\end{center}
\end{figure}

In the figures \ref{FIG1} and \ref{FIG2} are presented projections  of the flow of \eqref{Example1} restricted to the invariant set $\{\sigma_5=0\}\subset \bar{\Psi}_\epsilon$ on the planes $(\sigma_2, v)$ and $(\sigma_2,\sigma_4)$, respectively, for the potential  $V(\phi)=V_1 e^{-\alpha \phi }+V_2 e^{-\beta\phi},$ and the coupling function $\chi=\chi_0 \exp\left[\frac{\lambda\phi}{4-3\gamma}\right]$.  We select the values of the parameters: $V_1=1, V_2=2, \alpha=1.5, \beta=2, \lambda=-5$ and $\gamma=1.$ These numerical simulations confirms the stability of $P_6$ for this choice of parameters.

\begin{figure}[t]
\begin{center}
\mbox{\epsfig{figure=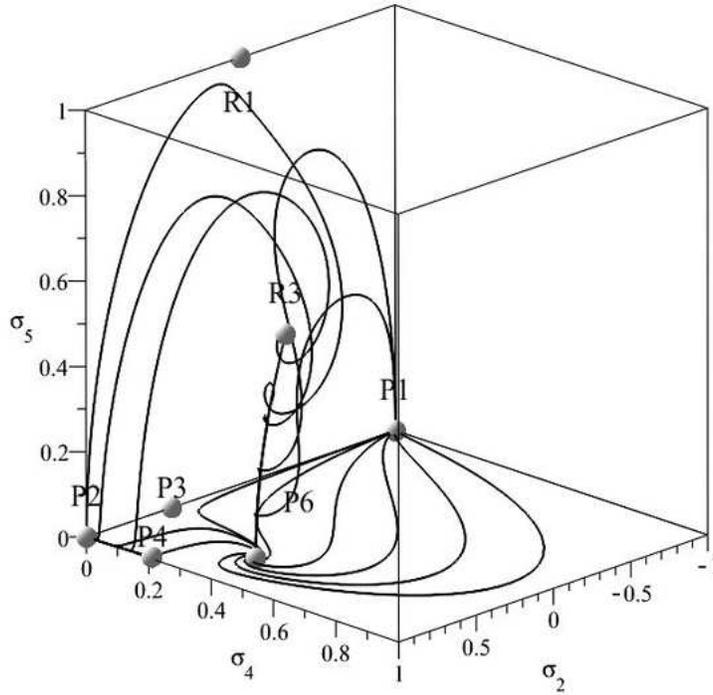,scale=0.5,angle=0}}
\caption{\label{FIG7}{Some orbits in the invariant set
$\sigma_2^2+\sigma_4^2+\sigma_5^2\leq 1$ for the flow of \eqref{Example1} restricted to the invariant set $\{v=0\}\subset \bar{\Psi}_\epsilon$ for the potential  $V(\phi)=V_1 e^{-\alpha \phi }+V_2 e^{-\beta\phi},$ and the coupling function $\chi=\chi_0 \exp\left[\frac{\lambda\phi}{4-3\gamma}\right]$.  We select the values of the parameters: $V_1=1, V_2=2, \alpha=2.4, \beta=3, \lambda=1.1$ and $\gamma=1.$ In the figure it is illustrated the stability of $P_6$. $P_1$ and $P_2$ are sources whereas $P_3, P_4, R_1$ and $R_3$ are saddles. $R_2$ does not exists.}} 
\end{center}
\end{figure}
In figure \ref{FIG7} we show some orbits in the invariant set
$\sigma_2^2+\sigma_4^2+\sigma_5^2\leq 1,\varphi=0$ for the potential  $V(\phi)=V_1 e^{-\alpha \phi }+V_2 e^{-\beta\phi},$ and the coupling function $\chi=\chi_0 \exp\left[\frac{\lambda\phi}{4-3\gamma}\right]$.  For the
choice $V_1=1, V_2=2, \alpha=2.4, \beta=3, \lambda=1.1$ and $\gamma=1$ results that powerlaw solution $P_6$ is the attractor. $P_1$ and $P_2$ are sources whereas $P_3, P_4, R_1$ and $R_3$ are saddles. $R_2$ does not exists.

\subsection{Coupling Functions and Potentials of Exponential Orders $M=0$ and $N=-\mu\neq 0,$ Respectively}

As an example let us consider $\chi,V\in {\cal E}^2_+$ of
exponential orders $M=0$ and $N=-\mu,$ respectively \cite{Nova_Science}. This class of
potentials contains the cases investigated in
\cite{Copeland:1997et,vandenHoogen:1999qq} (there are not
considered coupling to matter, i.e., $\chi(\phi)\equiv 1$, in the
second case, for flat FLRW cosmologies), the case investigated in
\cite{Copeland:2009be} (for positive potentials and standard flat FLRW
dynamics), the example examined in \cite{Leon:2008de}, etc. 

In table \ref{critcrit} are summarized the location, existence conditions
and stability of the singular points for the flow at the invariant set $\varphi=0$.

\begin{table}[!t]
\caption{\label{critcrit} Location of the singular points of the flow
of \eqref{Syst:26} defined in the invariant set
$\left\{p\in\Omega_\epsilon: \varphi=0\right\}$ for $M=0$ and
$N=-\mu.$ (Taken from \cite{Nova_Science}).}\bigskip 
\resizebox{1.0\textwidth}{!}{\begin{tabular}[t]{|l|c|c|c|c|}
\hline
Label&$(\sigma_2,\sigma_4,\sigma_5)$&Existence&Stability$^{\rm a}$\\[2ex]
\hline
$P_1$&$(-1,0,0)$&always & unstable if $\mu>-\sqrt{6}$\\[2ex]
&&& saddle otherwise \\
\hline
$P_2$&$(1,0,0)$&always & unstable if $\mu<\sqrt{6}$\\[2ex]
&&& saddle otherwise \\
\hline
$P_3$&$(0,0,0)$&always & saddle\\[2ex]
\hline
$R_1$&$(0,0,1)$&always & saddle\\[2ex]
\hline
$P_4$&$\left(\frac{\mu}{\sqrt{6}},\sqrt{1-\frac{\mu^2}{6}},0\right)$&$\mu^2<6$ & stable for $\left\{\begin{array}{c} 0<\gamma<\frac{4}{3},\,\,\mu^2<3\gamma,\, \text{or}\\
\frac{4}{3}<\gamma<2,\,\, \mu^2<2
\end{array},\right.$\\
&&& saddle otherwise\\[2ex]
\hline
$P_{5,6}$&$\left(\sqrt{\frac{3}{2}}\frac{\gamma}{\mu}, \pm \frac{1}{\mu}\sqrt{\frac{3}{2}(2-\gamma)\gamma},0 \right)$&$\mu^2>3\gamma$ & stable for $\left\{\begin{array}{c} 0<\gamma<\frac{2}{9},\,\mu^2>3\gamma,\, \text{or}\\
\frac{2}{9}<\gamma<\frac{4}{3},\,\,
3\gamma<\mu^2<\frac{24\gamma^2}{9\gamma-2},\, \text{or}\\
\frac{2}{9}<\gamma <\frac{4}{3},\,\mu ^2>\frac{24 \gamma ^2}{9
\gamma -2}
\end{array},\right.$\\
&&& saddle otherwise\\[2ex]
\hline
$R_3$ & $\left(\frac{2\sqrt{\frac{2}{3}}}{\mu},\frac{2}{\sqrt{3}|\mu|},\frac{\sqrt{\mu^2-4}}{|\mu|}\right)$& $|\mu|>2$ & stable if $\frac{4}{3}<\gamma<2,$ saddle otherwise \\[2ex]
\hline
\end{tabular}}
$^{\rm a}$ The stability is analyzed for the flow restricted to
the invariant set $\varphi=0$.
\end{table}

Let us discuss some physical properties of the cosmological
solutions associated to the singular points displayed in table
\ref{critcrit} \cite{Nova_Science}:
\begin{itemize}
\item  $P_{1,2}$  represent kinetic-dominated cosmological
solutions. They behave as stiff-like matter and can be local past attractors of the systems for an open set of values of the parameter $\mu.$
The same rates for $a, H,$ and $\phi$ presented in section \eqref{Sect:double_exp} applies here too. 
\item  $P_3$ represents matter-dominated cosmological solutions
that satisfy $H=\frac{2}{3 t \gamma -2 c_1},a=\left(3 t \gamma -2
c_1\right){}^{\frac{2}{3 \gamma }} c_2,\rho_m
   =\frac{12}{\left(3 t \gamma -2 c_1\right){}^2}+c_3.$
\item $R_1$ represents a radiation-dominated cosmological
solutions satisfying $H=\frac{1}{2 t-c_1},a=\sqrt{2 t-c_1}
c_2,\rho_r=\frac{3}{\left(2 t-c_1\right){}^2}+c_3.$ \item $P_4$
represents power-law scalar-field dominated inflationary
cosmological solutions. As $\phi\rightarrow +\infty$ the potential
behaves as $V\sim V_0 \exp[-\mu \phi]$. Thus it is easy to obtain
the asymptotic exact solution: $H=\frac{2}{t \mu ^2-2
c_1},a=\left(t \mu ^2-2 c_1\right){}^{\frac{2}{\mu ^2}} c_2,\phi
\sim\frac{1}{\mu}\ln\left[\frac{V_0(t\mu^2-2c_1)^2}{2(6-\mu^2)}\right].$
\item $P_{5,6}$ represent matter-kinetic-potential scaling
solutions. As before, in the limit $\phi\rightarrow +\infty$ we obtain
the asymptotic expansions: $H=\frac{2}{3 t \gamma -2
c_1},a=\left(3 t \gamma -2 c_1\right){}^{\frac{2}{3 \gamma }}
c_2,\phi \sim \frac{1}{\mu}\ln\left[\frac{V_0\mu^2(3t\gamma-2
c_1)^2}{18(2-\gamma)\gamma}\right].$ \item $R_3$ represent
radiation-kinetic-potential scaling solutions. As before are
deduced the following asymptotic expansions: $H=\frac{1}{2
t-c_1},a=\sqrt{2 t-c_1} c_2,\phi \sim
\frac{1}{\mu}\ln\left[\frac{v_0\mu^2(2t-c_1)^2}{4}\right].$
\end{itemize}

\subsubsection{Powerlaw coupling and Albrecht-Skordis
potential.}\label{toy}

Now, let's revisit the example of a powerlaw coupling function and an Albrecht-Skordis potential, first introduced in \cite{Leon:2008de}, and then extended in section 4.4 of \cite{Nova_Science}.

Let us consider the coupling function \be
\chi(\phi)=\left(\frac{3\alpha}{8}\right)^{\frac{1}{\alpha}}\chi_0(\phi-\phi_0)^
{\frac{2}{\alpha}},\; \alpha>0,\text{const.},\,\phi_0\geq
0.\label{couplingexample}\ee and the Albrecht-Skordis  potential \cite{Albrecht:1999rm}:
\begin{equation}
V(\phi )=e^{-\mu \phi }{\left( A+(\phi -B)^2\right).}
\label{Albrecht-Skordis}
\end{equation}

The coupling function (\ref{couplingexample}) and the
potential (\ref{Albrecht-Skordis}) are WBI of exponential orders
$M=0$ and $N=-\mu,$ respectively.

It is easy to prove that Power-law coupling and the
Albrecht-Skordis potential are at least  ${\cal E}^2_+,$  under
the admissible coordinate transformation \cite{Nova_Science}: \footnote{We fix here an
error in formulas B6-B9 in \cite{Leon:2008de}. With the choice
$\vphi=\phi^{-1}$ the resulting barred functions given by B7-B9
there, are not of the desired differentiable class.}
\be \vphi=\phi^{-\frac{1}{2}}=f(\phi)\label{transformAS}.\ee

Using the above coordinate transformation we find
\be\overline{W}_{\chi}(\vphi)=\frac{2 \varphi^2 }{\alpha  (1-\varphi^2  \phi_0)}.\label{WchiAS2}\ee
\be\overline{W}_V(\vphi)=-\frac{2\vphi^2(B\vphi^2-1)}{A\vphi^4+(B\vphi^2-1)^2}.\label{WVAS2}\ee
and
\be\overline{f'}(\vphi)=-\frac{1}{2}\vphi^3.\label{fAS2}\ee

In this example, the evolution equations for $\varphi,$
$\sigma_2,$ $\sigma_4,$ and $\sigma_5$ are given by the equations
\eqref{Syst:26} with $M=0,\,N=-\mu,$ and
$\overline{W}_{\chi}(\varphi),\,\overline{W}_{V}(\varphi)=0,$ and
$\overline{f'},$ given by \eqref{WchiAS2}, \eqref{WVAS2} and
\eqref{fAS2} respectively. The state space is defined by
$$\Omega_\epsilon=\{(\varphi, \sigma_2,
\sigma_4,\sigma_5)\in\mathbb{R}^4: 0\leq\varphi\leq
 \sqrt{\epsilon}, \sigma_2^2+\sigma_4^2+\sigma_5^2\leq 1,
\sigma_4\geq 0, \sigma_5\geq 0\}.$$

Finally, let us discuss some numerical simulations.
\begin{figure}[t]
\begin{center}
\hspace{0.8cm} \put(195,-3){${P_4}$} \put(215,-3){${P_2}$}
\put(0,-3){${P_1}$}\put(107,205){${\varphi}$}\put(235,3){${\sigma_2}$}
\includegraphics[width=8cm, height=7cm]{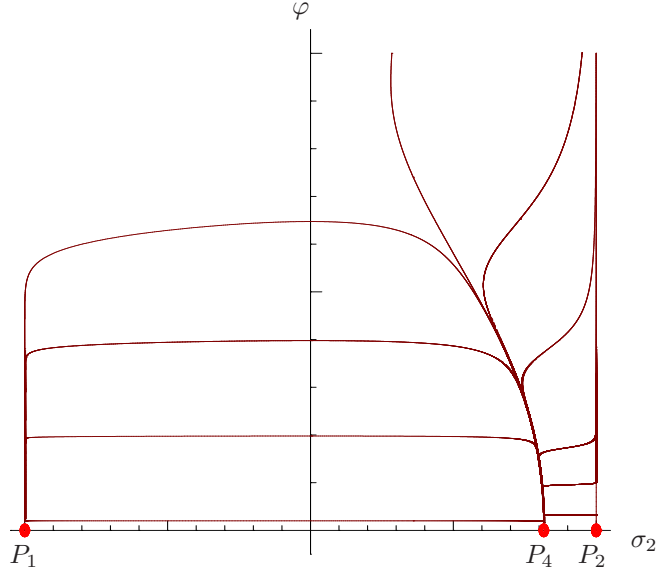}
\caption[Orbits in the invariant set  $\{\sigma_3=0\}\subset
\bar{\Sigma}_\epsilon$ for the coupling function
(\ref{couplingexample}) and potential
(\ref{Albrecht-Skordis}).]{Orbits in the invariant set
$\{\sigma_3=0,\sigma_5=0\}\subset \bar{\Sigma}_\epsilon$ for the model with
coupling function (\ref{couplingexample})  potential
(\ref{Albrecht-Skordis}). We select the values of the parameters:
$\epsilon=1.00,$ $\mu= 2.00, A = 0.50, \alpha = 0.33, B = 0.5,$
 $\phi_0=0,$ and $\gamma=1.$ Observe that i) almost all the orbits are past
asymptotic to $P_1;$ ii) $P_2$ is a saddle, and iii) the center
manifold of $P_4$ attracts all the orbits in the $\{\sigma_3=0\}$.
However, it is not an attractor in the invariant set
$\sigma_3>0,\,\varphi=0$ (see figure \ref{FIG6}). (Taken from \cite{Lap_Lambert}).} \label{FIG5}
\end{center}
\end{figure}

\begin{figure}[t]
\begin{center}
\hspace{0.4cm} \put(195,-3){${P_4}$} \put(215,-3){${P_2}$}
\put(0,-3){${P_1}$}\put(180,105){${P_5}$} \put(100,175){${P_3}$}
\put(107,205){${\sigma_3}$}\put(235,3){${\sigma_2}$}
\includegraphics[width=8cm, height=7cm]{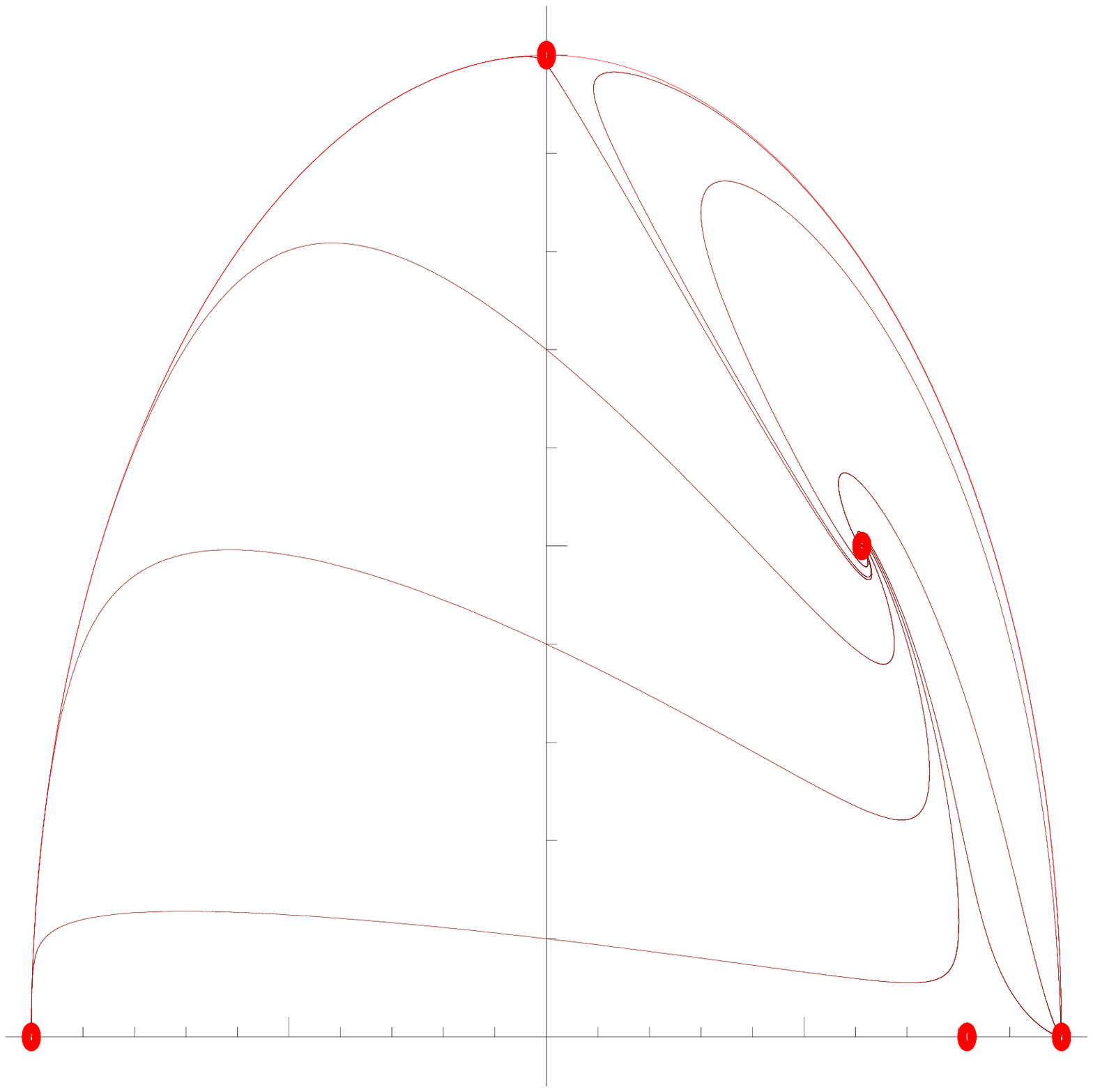}
\caption[Orbits in the invariant set $\{\varphi=0,\sigma_5=0\}\subset
\bar{\Sigma}_\epsilon$ for the coupling function
(\ref{couplingexample}) and potential
(\ref{Albrecht-Skordis}).]{Orbits in the invariant set
$\{\varphi=0, \sigma_5=0\}\subset \bar{\Sigma}_\epsilon$ for the model with
coupling function (\ref{couplingexample}) and potential
(\ref{Albrecht-Skordis}). We select the values of the parameters:
$\epsilon=1.00,$ $\mu= 2.00, A = 0.50, \alpha = 0.33, B = 0.5,$
 $\phi_0=0,$ and $\gamma=1.$ In the figure i) $P_{1,2}$ are local past
attractors, but $P_1$ is the global past attractor; ii) $P_{3,4}$
are saddles, and iii) $P_5$ is a local future
attractor. (Taken from \cite{Lap_Lambert}).}\label{FIG6}
\end{center}
\end{figure}

In the figure \ref{FIG5} are presented some orbits in the invariant set
$\{\sigma_3=0,\sigma_5=0\}\subset \bar{\Sigma}_\epsilon$ for the model with
coupling function (\ref{couplingexample})  potential
(\ref{Albrecht-Skordis}). We select the values of the parameters:
$\epsilon=1.00,$ $\mu= 2.00, A = 0.50, \alpha = 0.33, B = 0.5,$
 $\phi_0=0,$ and $\gamma=1.$ Observe that almost all the orbits are past
asymptotic to $P_1;$  $P_2$ is a saddle, and the center
manifold of $P_4$ attracts all the orbits in the $\{\sigma_3=0\}$.
However, it is not an attractor in the invariant set
$\sigma_3>0,\,\varphi=0$. In figure \ref{FIG6} are presented some orbits in the invariant set
$\{\varphi=0, \sigma_5=0\}\subset \bar{\Sigma}_\epsilon$ for the same values of the parameters as before. $P_{1,2}$ are local past
attractors, but $P_1$ is the global past attractor;  $P_{3,4}$
are saddles, and  $P_5$ is a local future
attractor.

\begin{figure}[t]
\begin{center}
\mbox{\epsfig{figure=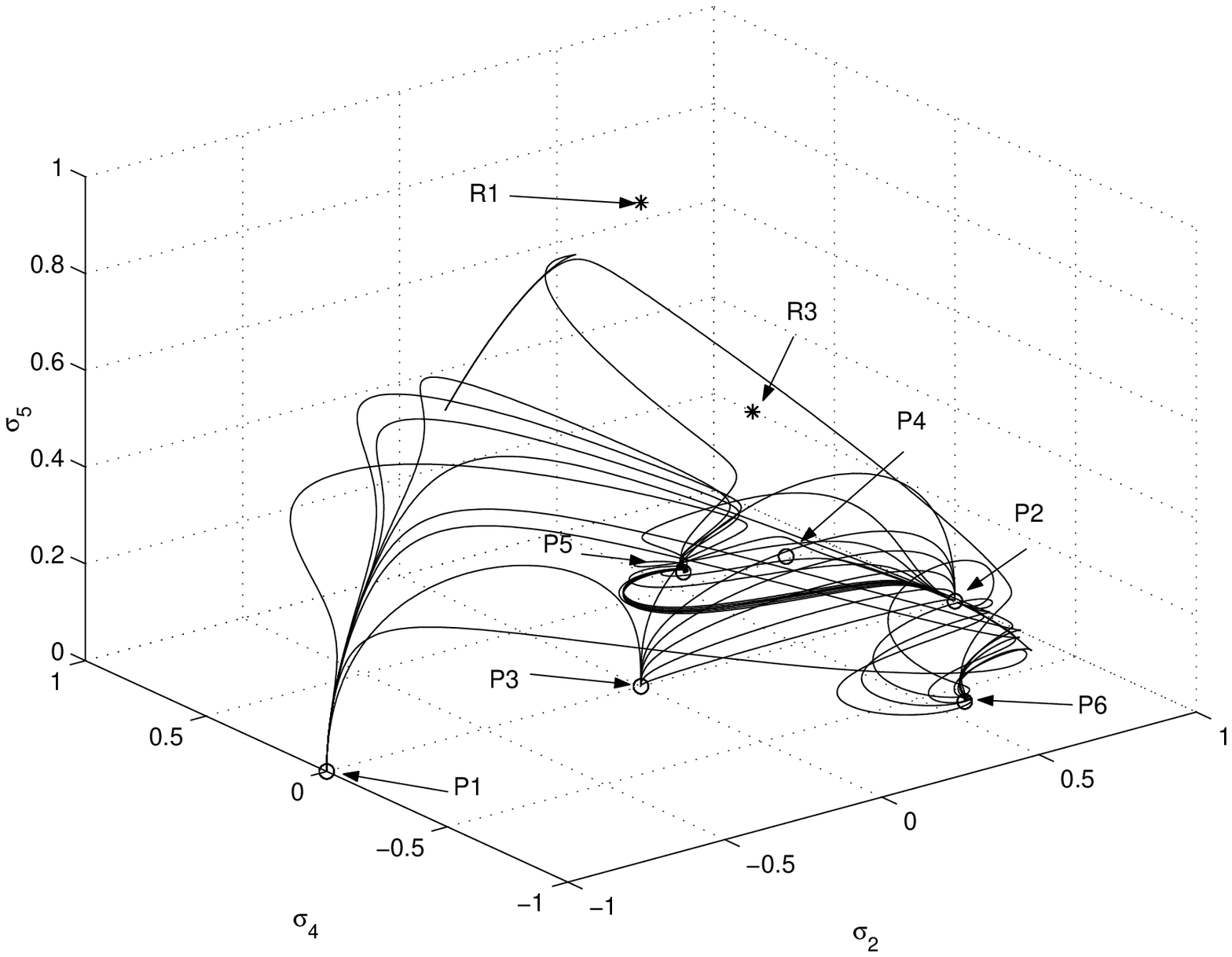,scale=0.6,angle=0}}
\caption{\label{FIG8}{Some orbits in the invariant set
$\sigma_2^2+\sigma_4^2+\sigma_5^2\leq 1$ for the choice of
$\varphi=0$ for the model with coupling function
(\ref{couplingexample})  potential (\ref{Albrecht-Skordis}). We
select the values of the parameters: $\gamma=1,$ $\epsilon=1.00,$
$\mu= 2.10, A = 0.50, \alpha = 0.33, B = 0.5,$ and $\phi_0=0.$ (Taken from \cite{Lap_Lambert}).}}
\end{center}
\end{figure}

In the figure \ref{FIG8} are displayed some orbits in the invariant set
$\sigma_2^2+\sigma_4^2+\sigma_5^2\leq 1$ for the choice of
$\varphi=0$ for the model with coupling function
(\ref{couplingexample})  potential (\ref{Albrecht-Skordis}). We
select the values of the parameters: $\gamma=1,$ $\epsilon=1.00,$
$\mu= 2.10, A = 0.50, \alpha = 0.33, B = 0.5,$ and $\phi_0=0.$  In this case $P_5$ is the local sink in this
invariant set.   $R_3$ exists and it is a saddle.

\section{Conclusion}

In this paper we have extended several results related to flat FLRW
models in the conformal (Einstein) frame of scalar-tensor gravity
theories. We have considered scalar fields with arbitrary (positive)
potentials and arbitrary coupling functions. Then, we have straightforwardly introduced mild assumptions under
such functions (differentiable class, number of singular points,
asymptotes, etc.) in order to characterize the asymptotic structure on a phase-space. Also, we have presented several numerical evidences that
confirm some of these results.

Our main results are the following.

\begin{enumerate}

\item Proposition \ref{Proposition I} states that for non-negative potentials with a
local zero minimum at $\phi=0;$ such that its derivative  is bounded in the
same set where the potential is; and provided the derivative of
the logarithm of the coupling function has an upper bound, then
the energy densities o matter and radiation as well as the kinetic term
tend to zero when the time goes forward. Thus, the Universe would expand
forever in a de Sitter phase in the future. This result is an extension of the
Proposition 2 in \cite{Miritzis:2003ym} to the non-minimal
coupling context. It is also an extension of Proposition 4 in \cite{Lap_Lambert} when the radiation is included in the cosmic budget.

\item Under the same hypotheses as in \ref{Proposition I} and provided that $V(\phi)$ is strictly decreasing
(increasing) for negative (positive) values of the scalar field,
then the scalar field diverges into the future or
it equals to zero (the last case holds only if the Hubble scalar
vanish towards the future). This Proposition \ref{Proposition II} is an extension of
proposition 3 in \cite{Miritzis:2003ym} and of Proposition 5 in \cite{Lap_Lambert} when the radiation is included in the background. 

\item Assuming that the potential is non-negative (with not
necessarily a local minimum at $(0,0)$), such that for $\phi\rightarrow +\infty$ it is unbounded. If its derivative is continuous
and bounded on a set $A$ where the potential is bounded. Then the cosmological model evolves to a late-time de
Sitter solution characterized by the divergence of
the scalar field. Additionally, if the potential vanish
asymptotically, the Hubble
scalar vanishes too (see Proposition \ref{Proposition III}). Proposition \ref{Proposition III}  is an extension for  $\rho_r>0$ of Proposition 6 discussed in \cite{Lap_Lambert}.

\item We have formulated and
proved the Proposition \ref{thmIII} (Proposition 3 in \cite{Nova_Science}) generalizing analogous result in
\cite{Giambo':2009cc}. Our result states that if the potential $V(\phi)$
is such that the (possibly empty set) where it is negative is
bounded and the (possibly empty) set of singular points of
$V(\phi)$ is finite, then, the singular point $${\bf
p}_*:=\left(\phi_*,y_*=0,{\rho_m}_*=0,\rho_r=0,
H=\sqrt{\frac{V(\phi_*)}{3}}\right),$$ where $\phi_*$ is a strict
local minimum for $V(\phi),$ corresponding to a \emph{de Sitter} solution,  is an asymptotically stable singular
point for the flow.

\item For the analysis of the system as $\phi\rightarrow\infty$ we
have defined a suitable change of variables to bring a
neighborhood of $\phi=\infty$ in a bounded set.  In this regime we found: radiation-dominated
cosmological solutions; power-law scalar-field dominated
inflationary cosmological solutions; matter-kinetic-potential
scaling solutions and radiation-kinetic-potential scaling
solutions. The rigorous mathematical apparatus was developed in section \ref{Sect:Inf}.

\item Using the above procedure we have investigated the behavior at the limit $\phi\rightarrow +\infty$ for the following models:  \\
(i)  a double exponential potential $V(\phi)=V_1 e^{-\alpha \phi }+V_2 e^{-\beta\phi}, \; 0<\alpha<\beta,$ and the coupling function $\chi=\chi_0 \exp\left[\frac{\lambda\phi}{4-3\gamma}\right]$, where $\lambda$ is a constant discussed in \cite{Tzanni:2014eja} and \\
(ii) a general class of potentials containing the
cases investigated in \cite{Copeland:1997et,vandenHoogen:1999qq} and in \cite{Albrecht:1999rm}.
We have re-examined the toy
model with power-law coupling and Albrecht-Skordis potential 
$V(\phi )=e^{-\mu \phi }{\left( A+(\phi -B)^2\right)}$ \cite{Albrecht:1999rm}
investigated in \cite{Leon:2008de} in presence of radiation.

\end{enumerate}

\section*{Acknowledgements}

CRF wish to thank the MES of Cuba for partial financial
support of this investigation. GL was supported by COMISI\'ON NACIONAL DE CIENCIAS Y
TECNOLOG\'IA through Proyecto FONDECYT DE POSTDOCTORADO 2014  grant  3140244
and by DI-PUCV grant 123.730/2013. The authors wish to thank to two anonymous referees for helpful suggestions.

\appendix

\section{Singular points of the flow of \eqref{Syst:26} in the phase space \eqref{PhaseSpace}.}
\label{AppendixA}

The system \eqref{Syst:26} admits the following
singular points (taken from \cite{Nova_Science}).

\begin{enumerate}
\item The singular point $P_1$ with coordinates
$\varphi=0,\sigma_2=-1,\sigma_4=0,\sigma_5=0$ always exists. The
eigenvalues of the linearisation around the singular point are $0,
\frac{1}{3},1-\frac{N}{\sqrt{6}},\frac{M (4-3 \gamma
)}{\sqrt{6}}-\gamma +2.$ Thus, 
\begin{enumerate}
\item $P_1$ has a 1-dimensional center manifold tangent to the $\varphi$-axis provided $N\neq\sqrt{6}$ and $M\neq -\frac{\sqrt{6} (\gamma -2)}{3 \gamma -4}$ (otherwise the center manifold would be 2- or 3-dimensional).
\item $P_1$ admits a 3-dimensional unstable manifold and a 1-dimensional center manifold for $N<\sqrt{6},\, 0<\gamma<\frac{4}{3},\,M>-\frac{\sqrt{6} (\gamma -2)}{3 \gamma -4};$ or
$N<\sqrt{6},\, \frac{4}{3}<\gamma<2,\,M<-\frac{\sqrt{6} (\gamma -2)}{3 \gamma -4}.$
In this case the center manifold of $P_1$ acts as a local source for an open set of orbits in \eqref{PhaseSpace}.
\item $P_1$ admits a 2-dimensional unstable manifold, a 1-dimensional stable manifold and a 1-dimensional center if
$N>\sqrt{6},\, 0<\gamma<\frac{4}{3},\,M>-\frac{\sqrt{6} (\gamma -2)}{3 \gamma -4};$ or
$N>\sqrt{6},\, \frac{4}{3}<\gamma<2,\,M<-\frac{\sqrt{6} (\gamma -2)}{3 \gamma -4};$ or
$N<\sqrt{6},\, 0<\gamma<\frac{4}{3},\,M<-\frac{\sqrt{6} (\gamma -2)}{3 \gamma -4};$ or
$N<\sqrt{6},\, \frac{4}{3}<\gamma<2,\,M>-\frac{\sqrt{6} (\gamma -2)}{3 \gamma -4}.$
\item $P_1$ admits a 1-dimensional unstable manifold, a 2-dimensional stable manifold and a 1-dimensional center manifold for $N>\sqrt{6},\, 0<\gamma<\frac{4}{3},\,M<-\frac{\sqrt{6} (\gamma -2)}{3 \gamma -4};$ or
$N>\sqrt{6},\, \frac{4}{3}<\gamma<2,\,M>-\frac{\sqrt{6} (\gamma -2)}{3 \gamma -4}.$
\end{enumerate}

\item The singular point $P_2$ with coordinates
$\varphi=0,\sigma_2=1,\sigma_4=0,\sigma_5=0$ always exists. The
eigenvalues of the linearization around the singular point are $0,
\frac{1}{3},1+\frac{N}{\sqrt{6}},-\gamma +\frac{M (3 \gamma
-4)}{\sqrt{6}}+2.$ As before, let us determine conditions on the
free parameters for the existence of center, unstable and stable
manifolds for $P_2$.
\begin{enumerate}
\item If $N\neq-\sqrt{6}$ and $M\neq \frac{\sqrt{6} (\gamma -2)}{3 \gamma -4}$ there exists a 1-dimensional center manifold tangent to the $\varphi$-axis, otherwise the center manifold would be 2- or 3-dimensional.
\item $P_2$ has a 3-dimensional unstable manifold a a 1-dimensional center manifold (tangent the $\varphi$-axis) if 
$N>-\sqrt{6},\,0<\gamma<\frac{4}{3},\,M<\frac{\sqrt{6} (\gamma -2)}{3 \gamma -4};$ or 
$N>-\sqrt{6},\,\frac{4}{3}<\gamma<2,\,M>\frac{\sqrt{6} (\gamma -2)}{3 \gamma -4}.$
In this case the center manifold of $P_2$ acts as a local source for an open set of orbits in \eqref{PhaseSpace}.
\item $P_2$ has a 2-dimensional unstable manifold a 1-dimensional stable and a 1-dimensional center manifold if
$N<-\sqrt{6},\,0<\gamma<\frac{4}{3},\,M<\frac{\sqrt{6} (\gamma -2)}{3 \gamma -4};$ or
$N<-\sqrt{6},\,\frac{4}{3}<\gamma<2,\,M>\frac{\sqrt{6} (\gamma -2)}{3 \gamma -4};$ or
$N>-\sqrt{6},\,0<\gamma<\frac{4}{3},\,M>\frac{\sqrt{6} (\gamma -2)}{3 \gamma -4};$ or
$N>-\sqrt{6},\,\frac{4}{3}<\gamma<2,\,M<\frac{\sqrt{6} (\gamma -2)}{3 \gamma -4}.$
\item $P_2$ has a 1-dimensional unstable manifold a 2-dimensional stable and a 1-dimensional center manifold if
$N<-\sqrt{6},\,0<\gamma<\frac{4}{3},\,M>\frac{\sqrt{6} (\gamma -2)}{3 \gamma -4};$ or
$N<-\sqrt{6},\,\frac{4}{3}<\gamma<2,\,M<\frac{\sqrt{6} (\gamma -2)}{3 \gamma -4}.$
\end{enumerate}
\item The singular point $P_3$ with coordinates
$\varphi=0,\sigma_2=\frac{M (3 \gamma -4)}{\sqrt{6} (\gamma
-2)},\sigma_4=0,\sigma_5=0$ exists for
\begin{enumerate}\item $0<\gamma <\frac{4}{3},\,-\frac{\sqrt{6} (\gamma -2)}{3 \gamma -4}\leq M\leq \frac{\sqrt{6} (\gamma -2)}{3 \gamma
   -4};$ or \item $\frac{4}{3}<\gamma <2,\,\frac{\sqrt{6} (\gamma -2)}{3 \gamma -4}\leq M\leq -\frac{\sqrt{6} (\gamma
   -2)}{3 \gamma -4}.$\end{enumerate}
The eigenvalues of the linearization are
$0,\,\lambda_1=-\frac{(3 \gamma -4) \left((3 \gamma -4) M^2-2 \gamma +4\right)}{12 (\gamma -2)},\,\lambda_2=\frac{-M^2 (4-3 \gamma )^2+6 (\gamma
   -2) \gamma +2 M N (3 \gamma -4)}{12 (\gamma -2)},\,\lambda_3=\frac{6 (\gamma -2)^2-M^2 (4-3 \gamma )^2}{12 (\gamma -2)}.$
As before, let us determine conditions on the free parameters for the existence of center, unstable and stable manifolds for $P_3$.
\begin{enumerate}
\item For $\gamma, N$ and $M$ such that $\lambda_{1}\neq 0, \lambda_{2}\neq 0, \lambda_{3}\neq 0$ the center manifold is 1-dimensional and tangent to the $\varphi$-axis. Otherwise the center manifold coud be 2-, or 3-dimensional (it is never 4-dimensional).
\item $P_3$ admits a 1-dimensional center manifold and a 3-dimensional stable manifold for
$0<\gamma<\frac{4}{3},\,-\frac{\sqrt{2} \sqrt{\gamma -2}}{\sqrt{3 \gamma -4}}<M<0,\,\ N>\frac{M^2 (4-3 \gamma
   )^2-6 (\gamma -2) \gamma }{2 M (3 \gamma -4)};$ or
$0<\gamma<\frac{4}{3},\,0<M<\frac{\sqrt{2} \sqrt{\gamma -2}}{\sqrt{3 \gamma -4}},\,
   N<\frac{M^2 (4-3 \gamma )^2-6 (\gamma -2) \gamma }{2 M (3 \gamma -4)}.$
\item The unstable manifold  of $P_3$ is 2-dimensional (thus its stable and center manifolds are both 1-dimensional) in the cases
$0<\gamma <\frac{4}{3},\,-\frac{\sqrt{6} (\gamma -2)}{3 \gamma -4}<M<-\frac{\sqrt{2} \sqrt{\gamma -2}}{\sqrt{3 \gamma -4}},N<\frac{M^2 (4-3 \gamma
   )^2-6 (\gamma -2) \gamma }{2 M (3 \gamma -4)};$ or
$0<\gamma <\frac{4}{3},\,\frac{\sqrt{2} \sqrt{\gamma -2}}{\sqrt{3 \gamma -4}}<M<\frac{\sqrt{6} (\gamma -2)}{3 \gamma -4},N>\frac{M^2 (4-3 \gamma
   )^2-6 (\gamma -2) \gamma }{2 M (3 \gamma -4)};$ or
$\frac{4}{3}<\gamma <2,\frac{\sqrt{6} (\gamma -2)}{3 \gamma -4}<M<0,N>\frac{M^2 (4-3 \gamma )^2-6 (\gamma -2) \gamma }{2 M
   (3 \gamma -4)};$ or
$\frac{4}{3}<\gamma <2, M=0, N\in\mathbb{R};$ or
$\frac{4}{3}<\gamma <2,0<M<-\frac{\sqrt{6} (\gamma -2)}{3 \gamma -4},N<\frac{M^2 (4-3 \gamma )^2-6 (\gamma -2) \gamma }{2 M
   (3 \gamma -4)},$ 
Otherwise, $P_3$ has a 1-dimensional unstable manifold. Thus, it is never a local source since its unstable manifold is of dimension less than $3.$
\end{enumerate}
\item The singular point $R_1$ with coordinates
$\varphi=0,\sigma_2=0,\sigma_4=0,\sigma_5=1$ always exists. The
eigenvalues of the linearization are
$0,\frac{2}{3},-\frac{1}{3},\frac{4}{3}-\gamma.$ The center
manifold is 1-dimensional and tangent to the $\varphi$-axis. The
unstable (stable) manifold is 1-dimensional (2-dimensional) if
$\gamma>\frac{4}{3}$ otherwise it is 2-dimensional
(1-dimensional). 
\item The singular point $R_2$ with coordinates
$\sigma_2=\frac{\sqrt{\frac{2}{3}}}{M},\sigma_4=0,\sigma_5=\frac{\sqrt{\frac{4-2
\gamma }{M^2}+3 \gamma -4}}{\sqrt{3 \gamma -4}}$ exists for
$0<\gamma<\frac{4}{3},\,M^2\geq \frac{2\left(\gamma -2\right)}{3
\gamma -4}.$ The eigenvalues of the linearization are
$0,-\frac{M+\sqrt{3 M^2 (4 \gamma -5)-8 (\gamma -2)}}{6 M},\frac{\sqrt{3 M^2 (4 \gamma -5)-8 (\gamma -2)}-M}{6
   M},\frac{1}{3} \left(\frac{N}{M}+2\right).$
Let us determine conditions on the free parameters for the existence of center, unstable and stable manifolds for $R_2$.
\begin{enumerate}
\item $R_2$ has a 3-dimensional stable manifold and a 1-dimensional center manifold if
$0<\gamma <\frac{5}{4},-2 \sqrt{\frac{2}{3}} \sqrt{\frac{\gamma -2}{4 \gamma -5}}\leq M<-\sqrt{2} \sqrt{\frac{\gamma -2}{3
   \gamma -4}},N>-2 M;$ or
$0<\gamma <\frac{5}{4},\sqrt{2} \sqrt{\frac{\gamma -2}{3 \gamma -4}}<M\leq 2 \sqrt{\frac{2}{3}} \sqrt{\frac{\gamma -2}{4
   \gamma -5}},N<-2 M;$ or
$\frac{5}{4}\leq \gamma <\frac{4}{3},M<-\sqrt{2} \sqrt{\frac{\gamma -2}{3 \gamma -4}},N>-2 M;$  or
$\frac{5}{4}\leq \gamma <\frac{4}{3},M>\sqrt{2} \sqrt{\frac{\gamma -2}{3 \gamma -4}},N<-2 M;$ or
$0<\gamma <\frac{5}{4},M<-2 \sqrt{\frac{2}{3}} \sqrt{\frac{\gamma -2}{4 \gamma -5}}, N>-2 M;$ or
$0<\gamma <\frac{5}{4},M>2 \sqrt{\frac{2}{3}} \sqrt{\frac{\gamma -2}{4 \gamma -5}}, N<-2 M.$

\item By reversing the sign of the last inequality, i.e., the inequality solved for $N$, in the previous six cases we obtain conditions for $R_2$ having a 2-dimensional stable manifold, a 1-dimensional unstable manifold and a 1-dimensional center manifold.
\end{enumerate}
\item The singular point $P_4$ with coordinates $\varphi=0,
\sigma_2=-\frac{N}{\sqrt{6}},\sigma_4=\sqrt{1-\frac{N^2}{6}},
\sigma_5=0$ exists whenever $N^2<6.$ The eigenvalues of the
linearization are
$0,\frac{1}{6} \left(N^2-6\right),\frac{1}{6} \left(N^2-4\right),\frac{1}{3} N (2 M+N)-\frac{1}{2} (M N+2) \gamma.$ The conditions for the existence of stable, unstable and center manifolds is as follows.
\begin{enumerate}
\item The center manifold is 1-dimensional and the stable manifold is 3-dimensional provided
 $N=0,\,M\in\mathbb{R},\gamma\neq \frac{4}{3};$ or
 $0<\gamma <\frac{4}{3},-2<N<0,M>\frac{2 \left(N^2-3 \gamma \right)}{N (3 \gamma -4)};$ or
 $0<\gamma <\frac{4}{3},0<N<2,M<\frac{2 \left(N^2-3 \gamma \right)}{N (3 \gamma -4)};$ or
 $\frac{4}{3}<\gamma <2,-2<N<0,M<\frac{2 \left(N^2-3 \gamma \right)}{N (3 \gamma -4)};$ or
 $\frac{4}{3}<\gamma <2,0<N<2,M>\frac{2 \left(N^2-3 \gamma \right)}{N (3 \gamma -4)}.$
\item The stable manifold is 2-dimensional, the unstable manifold is 1-dimensional and the center manifold is 1-dimensional provided
 $0<\gamma <\frac{4}{3},-\sqrt{6}<N<-2,M>\frac{2 \left(N^2-3 \gamma \right)}{N (3 \gamma -4)};$ or
 $0<\gamma <\frac{4}{3},-2<N<0,M<\frac{2 \left(N^2-3 \gamma \right)}{N (3 \gamma -4)};$ or
 $0<\gamma <\frac{4}{3},0<N<2,M>\frac{2 \left(N^2-3 \gamma \right)}{N (3 \gamma -4)};$ or
 $0<\gamma <\frac{4}{3},2<N<\sqrt{6},M<\frac{2 \left(N^2-3 \gamma \right)}{N (3 \gamma -4)};$ or
 $\frac{4}{3}<\gamma <2,-\sqrt{6}<N<-2,M<\frac{2 \left(N^2-3 \gamma \right)}{N (3 \gamma -4)};$ or
 $\frac{4}{3}<\gamma <2,-2<N<0,M>\frac{2 \left(N^2-3 \gamma \right)}{N (3 \gamma -4)};$ or
 $\frac{4}{3}<\gamma <2,0<N<2,M<\frac{2 \left(N^2-3 \gamma \right)}{N (3 \gamma -4)};$ or
 $\frac{4}{3}<\gamma <2,2<N<\sqrt{6},M>\frac{2 \left(N^2-3 \gamma \right)}{N (3 \gamma -4)}.$

\item The stable manifold is 1-dimensional, the unstable manifold is 2-dimensional and the center manifold is 1-dimensional provided
 $0<\gamma <\frac{4}{3},-\sqrt{6}<N<-2,M<\frac{2 \left(N^2-3 \gamma \right)}{N (3 \gamma -4)};$ or
 $0<\gamma <\frac{4}{3},2<N<\sqrt{6},M>\frac{2 \left(N^2-3 \gamma \right)}{N (3 \gamma -4)};$ or
 $\frac{4}{3}<\gamma <2,-\sqrt{6}<N<-2,M>\frac{2 \left(N^2-3 \gamma \right)}{N (3 \gamma -4)};$ or
 $\frac{4}{3}<\gamma <2,2<N<\sqrt{6},M<\frac{2 \left(N^2-3 \gamma \right)}{N (3 \gamma -4)}.$
\end{enumerate}
\item The singular point $R_3$ with coordinates
$\varphi=0,\sigma_2=-\frac{2
\sqrt{\frac{2}{3}}}{N},\sigma_4=\frac{2}{\sqrt{3}
|N|},\sigma_5=\frac{\sqrt{N^2-4}}{|N|}$ exists for $N^2\geq 4.$
The eigenvalues of the linearization are
$0,\frac{1}{6} \left(-\frac{\sqrt{64 N^2-15 N^4}}{N^2}-1\right),\frac{1}{6} \left(\frac{\sqrt{64 N^2-15
   N^4}}{N^2}-1\right),-\frac{(2 M+N) (3 \gamma -4)}{3 N}.$
The conditions for the existence of stable, unstable and center manifolds are as follows.
\begin{enumerate}
\item The stable manifold is 3-dimensional and the center manifold is 1-dimensional provided
 $0<\gamma <\frac{4}{3},N<-\frac{8}{\sqrt{15}}, M>-\frac{N}{2};$ or
 $0<\gamma <\frac{4}{3},-\frac{8}{\sqrt{15}}\leq N<-2,M>-\frac{N}{2};$ or
 $0<\gamma <\frac{4}{3},2<N\leq \frac{8}{\sqrt{15}},M<-\frac{N}{2};$ or
 $0<\gamma <\frac{4}{3},N>\frac{8}{\sqrt{15}},M<-\frac{N}{2};$ or
 $\frac{4}{3}<\gamma <2,N<-\frac{8}{\sqrt{15}},M<-\frac{N}{2};$ or
 $\frac{4}{3}<\gamma <2,-\frac{8}{\sqrt{15}}\leq N<-2,M<-\frac{N}{2};$ or
 $\frac{4}{3}<\gamma <2,2<N\leq \frac{8}{\sqrt{15}},M>-\frac{N}{2};$ or
 $\frac{4}{3}<\gamma <2,N>\frac{8}{\sqrt{15}},M>-\frac{N}{2}.$

\item By reversing the sign of the last inequality, i.e., the inequality solved for $M$, in the previous eight cases we obtain conditions for $R_3$ having a 2-dimensional stable manifold, a 1-dimensional unstable manifold and a 1-dimensional center manifold.
\end{enumerate}
\item The singular point $P_5$ with coordinates

$\varphi=0,\sigma_2=\frac{\sqrt{6} \gamma }{M (3 \gamma -4)-2 N},$ $\sigma_4=\frac{\sqrt{M^2 (4-3 \gamma )^2+M N (8-6 \gamma )-6 (\gamma -2) \gamma }}{2 N+M (4-3 \gamma )},\sigma_5=0$ exists for $2 (2 M+N)>3 M \gamma,\, M (3 \gamma -4) (M (3 \gamma -4)-2 N)\geq 6 (\gamma -2) \gamma,$
and
$\frac{3 (M N+2) \gamma -2 N (2
   M+N)}{(2 N+M (4-3 \gamma ))^2}\leq 0.$ The eigenvalues of the linearization are

$0,\frac{12 M+6 N-3 (3 M+N) \gamma +\sqrt{3} \sqrt{f(\gamma ,M,N)}}{6 (M (3 \gamma -4)-2 N)},\frac{3 N (\gamma -2)+3 M (3
   \gamma -4)+\sqrt{3} \sqrt{f(\gamma ,M,N)}}{6 (2 N+M (4-3 \gamma ))},\frac{(2 M+N) (3 \gamma -4)}{6 N+3 M (4-3 \gamma
   )},$
   where $f(\gamma ,M,N)=2 M^3 N (3 \gamma -4)^3+2 M N \left(4 N^2-6 \gamma ^2+3 \gamma -6\right) (3 \gamma -4)-M^2 \left(8 N^2-12 \gamma
   -3\right) (4-3 \gamma )^2+3 (\gamma -2) \left(N^2 (9 \gamma -2)-24 \gamma ^2\right).$ The stability conditions of $P_5$ are very complicated to display them here. Thus we must rely on numerical experimentation. We can obtain, however, some analytic results. For instance, there exists at least a 1-dimensional center manifold. The unstable manifold is always of dimension lower than 3. Thus the singular point is never a local source. If all the eigenvalues, apart form the zero one, have negative reals parts, then the center manifold of $P_5$ acts as a local sink. This means that the orbits in the stable manifold approach the center manifold of $P_5$ when the time goes forward.
\item The singular point $P_6$ with coordinates

$\varphi=0,\sigma_2=\frac{\sqrt{6} \gamma }{M (3 \gamma -4)-2 N},$ $\sigma_4=-\frac{\sqrt{M^2 (4-3 \gamma )^2+M N (8-6 \gamma )-6 (\gamma -2) \gamma }}{2 N+M (4-3 \gamma )},\sigma_5=0$ exists for $M (3 \gamma -4) (M (3 \gamma -4)-2 N)\geq 6 (\gamma -2) \gamma ,\,2 (2 M+N)<3 M \gamma,$
and
$\frac{3 (M N+2) \gamma -2 N (2
   M+N)}{(2 N+M (4-3 \gamma ))^2}\leq 0.$ The eigenvalues of the linearization are the same displayed in the previous point. However the stability conditions are rather different (since the existence conditions are different from those of $P_5$). As before, the stability conditions are very complicated to display them here, but similar conclusions concerning the center and unstable manifold, as for $P_5,$ are obtained. For get further information about its stability we must to resort to numerical experimentation.
\end{enumerate}

\section{Physical description of the solutions and connection with observables.}\label{AppendixB}

Let us now present the formalism of obtaining the physical
description of a singular point, and also connect with the basic
observables relevant for a physical discussion (taken from \cite{Nova_Science}).

Firstly, around a singular point we obtain first-order expansions
for $H,a,\phi,$ and $\rho_m$ and $\rho_r$ in terms of $t$,
considering equations: \eqref{Raych}; the definition of the scale
factor $a$ in terms of the Hubble factor $H$; the definition of
$\sigma_2;$  the matter conservation equations \eqref{consm} and
\eqref{consr}, respectively, given by
\begin{align}
&& 2 \dot H(t)=H(t)^2 \left(3 (\gamma -2) {\sigma_2^\star}^2+3 \gamma  \left({\sigma_4^\star}^2+{\sigma_5^\star}^2-1\right)-4
   {\sigma_5^\star}^2\right),\nonumber\\&& \dot a(t)=a(t) H(t),\nonumber\\&& \dot\phi(t)=\sqrt{6} {\sigma_2^\star} H(t),\nonumber\\&& \dot\rho_m(t)=-\frac{3}{2} H(t)^3
   \left(\sqrt{6} M (3 \gamma -4) {\sigma_2^\star}-6 \gamma \right) \left({\sigma_2^\star}^2+{\sigma_4^\star}^2+{\sigma_5^\star}^2-1\right),\nonumber\\&& \dot\rho_r(t)=-12 {\sigma_5^\star}^2 H(t)^3,\label{APPROX}
\end{align}
where the star-superscript denotes the evaluation at a specific
singular point. The equation \be \ddot \phi(t)= \frac{3}{2} H(t)^2
\left(M (3 \gamma -4)
\left({\sigma_2^\star}^2+{\sigma_4^\star}^2+{\sigma_5^\star}^2-1\right)-2
\left(N {\sigma_4^\star}^2+\sqrt{6}
{\sigma_2^\star}\right)\right),\label{consistency}\ee derived from
the equation of motion for the scalar field \eqref{KG} should be
used as a consistency test for the above procedure.  Solving the
differential equations \eqref{APPROX} and substituting the
resulting expressions in the equation \eqref{consistency} results
in
\begin{align}-6 M (3 \gamma -4) \left({\sigma_2^\star}^2+{\sigma_4^\star}^2+{\sigma_5^\star}^2-1\right)+12 N {\sigma_4^\star}^2+\nonumber\\+2 \sqrt{6}
   {\sigma_2^\star} \left(3 \gamma  \left({\sigma_2^\star}^2+{\sigma_4^\star}^2+{\sigma_5^\star}^2-1\right)-6 {\sigma_2^\star}^2-4 {\sigma_5^\star}^2+6\right)=0.\end{align}

This integrability condition should be (at least asymptotically) fulfilled.

\begin{table*}[t]
\caption{\label{tab2b} Observable cosmological quantities, and
physical behavior of the solutions, at the singular points of the
cosmological system. We use the notations
$M_1(\gamma)=\frac{\sqrt{2 \gamma  (3 \gamma -8)+8}}{4-3 \gamma
},$ $M_2(\gamma)=\frac{\sqrt{6} \sqrt{(\gamma -3) \gamma +2}}{4-3
\gamma }.$ (Taken from \cite{Nova_Science}).
 }\bigskip   \centering
\resizebox{1.0\textwidth}{!}
{\begin{tabular}{|c|c|c|c|}
  \hline   \hline
  \ \ Cr.P.  \ \  &\  $q$ \ & $w_{\text{eff}}$   & Solution/description\\
  \hline \hline
$P_1$ & 2 & 1 &  Decelerating. \\
   \hline
$P_2$ & 2 & 1 & Decelerating. \\
     \hline
$P_3$ & $\frac{-M^2 (4-3 \gamma )^2+2 \gamma  (3 \gamma -8)+8}{4 (\gamma -2)}$ & $-\frac{M^2 (4-3 \gamma
   )^2}{6 (\gamma -2)}+\gamma -1$ &  Accelerating for   \\
      &  & &   $0<\gamma <\frac{2}{3},\, -M_1(\gamma)<M<M_1(\gamma)$ \\ \hline 
$P_4$ & $\frac{1}{2} \left(N^2-2\right)$& $\frac{1}{3}
   \left(N^2-3\right)$ &  Accelerating for   \\
   &  & &   $-\sqrt{2}<N<\sqrt{2}$ \\
   &  & &   powerlaw-inflationary \\ \hline
$P_5$ & $\frac{3 (M+N) \gamma -2 (2 M+N)}{2 N+M (4-3 \gamma )}$ & $\frac{M
   (4-3 \gamma )-2 N (\gamma -1)}{M (3 \gamma -4)-2 N}$ &  Accelerating for   \\
   &  & & $\frac{3 (M+N) \gamma -2 (2 M+N)}{2 N+M (4-3 \gamma )}<0$   \\
   &  & &  matter-kinetic-potential scaling   \\ \hline 
$P_6$ & $\frac{3 (M+N) \gamma -2 (2 M+N)}{2 N+M (4-3 \gamma )}$ & $\frac{M
   (4-3 \gamma )-2 N (\gamma -1)}{M (3 \gamma -4)-2 N}$ &  Accelerating for   \\
   &  & & $\frac{3 (M+N) \gamma -2 (2 M+N)}{2 N+M (4-3 \gamma )}<0$   \\
   &  & &  matter-kinetic-potential scaling   \\ \hline 
   $R_1$ & 1 & $\frac{1}{3}$ & Decelerating. Radiation-dominated. \\
     \hline
$R_2$ & 1 & $\frac{1}{3}$ & Decelerating. \\
   &  & &  radiation-kinetic-potential scaling \\
     \hline
$R_3$ & 1 & $\frac{1}{3}$ & Decelerating. \\
   &  & &   radiation-kinetic-potential scaling. \\
     \hline
\end{tabular}}

\end{table*}

Instead of apply this procedure to a generic singular point here,
we submit the reader to section \ref{applications} for some worked
examples where this procedure has been applied. However we will
discuss on some cosmological observables.

We can calculate the deceleration parameter $q$ defined as usual as \cite{WE}
\begin{equation}
\label{qq}q=-\frac{a \ddot a}{a^2}.
\end{equation}
Additionally, we can calculate the effective (total) equation-of-state parameter of the universe $w_{\text{eff}}$, defined conventionally as
\begin{equation}
\label{weff}
w_{\text{eff}}\equiv\frac{p_{\text{tot}}}{\rho_{\text{tot}}},
\end{equation}
where $p_{\text{tot}}$ and $\rho_{\text{tot}}$ are respectively the
total isotropic pressure and the total energy density. Therefore, in terms of the auxiliary variables we have
\begin{eqnarray}
\label{qq2}
q&=& -\frac{3}{2} (\gamma -2) \sigma_2^2-\frac{3 \gamma  \sigma_4^2}{2}+\frac{1}{2} (4-3
   \gamma ) \sigma_5^2+\frac{1}{2} (3 \gamma -2)\\
w_{\text{eff}}&=&(2-\gamma ) \sigma_2^2-\gamma  \sigma_4^2+\frac{1}{3} (4-3 \gamma ) \sigma_5^2+\gamma -1. \label{weff2}
\end{eqnarray}

First of all, for each singular point described in the last
section we calculate the effective (total) equation-of-state
parameter of the universe $w_{\text{eff}}$ using \eqref{weff2},
and the deceleration parameter $q$ using \eqref{qq2}. The results
are presented in Table \ref{tab2b}. Furthermore, as usual, for an
expanding universe $q<0$ corresponds to accelerating expansion and
$q>0$ to decelerating expansion.

\label{lastpage-01}

\end{document}